**Engineered MoSe$_2$-based heterostructures for efficient electrochemical hydrogen evolution reaction**

*Leyla Najafi, Sebastiano Bellani, Reinier Oropesa-Nuñez, Alberto Ansaldo, Mirko Prato, Antonio Esau Del Rio Castillo and Francesco Bonaccorso*[*]

L. Najafi, Dr. S. Bellani, Dr. R. Oropesa-Nuñez, Dr. A. Ansaldo, Dr. A. E. Del Rio Castillo, Dr. F. Bonaccorso
Graphene Labs, Istituto Italiano di Tecnologia, via Morego 30, 16163 Genova, Italy
Università degli studi di Genova, Dipartimento di Chimica e Chimica Industriale, Via Dodecaneso 31, 16163 Genova, Italy
E-mail: francesco.bonaccorso@iit.it
Dr. M. Prato
Materials Characterization Facility, Istituto Italiano di Tecnologia, via Morego 30, 16163 Genova, Italy

Two-dimensional transition metal-dichalcogenides are emerging as efficient and cost-effective electrocatalysts for hydrogen evolution reaction (HER). However, only the edge sites of their trigonal prismatic phase show HER-electrocatalytic properties, while the basal plane, which is absent of defective/unsaturated sites, is inactive. Here, we tackle the key challenge that is increasing the number of electrocatalytic sites by designing and engineering heterostructures composed of single-/few-layer MoSe$_2$ flakes and carbon nanomaterials (graphene or single-wall carbon nanotubes (SWNTs)) produced by solution processing. The electrochemical coupling between the materials that comprise the heterostructure effectively enhances the HER-electrocatalytic activity of the native MoSe$_2$ flakes. The optimization of the mass loading of MoSe$_2$ flakes and their electrode assembly *via* monolithic heterostructure stacking provided a cathodic current density of 10mAcm$^{-2}$ at overpotential of 100mV, a Tafel slope of 63mVdec$^{-1}$ and an exchange current density ($j_0$) of 0.203μAcm$^{-2}$. In addition, electrode thermal annealing in a hydrogen environment and chemical bathing in n-butyllithium are exploited to texturize the basal planes of the MoSe$_2$ flakes (through Se-vacancies creation) and to achieve *in situ* semiconducting-to-metallic phase conversion, respectively, thus they activate new HER-electrocatalytic sites. The as-engineered electrodes show a 4.8-fold enhancement of $j_0$ and a decrease in the Tafel slope to 54mVdec$^{-1}$.

# 1. Introduction

Molecular hydrogen ($H_2$) that is produced from electrochemical water splitting has attracted a growing attention due to its high energy density (between 120-140 MJ kg$^{-1}$) and environmental friendliness.[1-3] The most effective $H_2$ evolution reaction (HER)-electrocatalysts are platinum-group elements[4] (*e.g.*, Pt[5,6] and Pd[7,8]), but their high cost (> 30 USD g$^{-1}$ for both Pt and Pd)[9] and scarcity (< 0.005 ppm of Pt and <0.001 ppm of Pd in Earth's crush)[10] hinder their use in mass commercial applications.[11] Therefore, HER-electrocatalysts based on Earth-abundant and electrochemically stable materials are being pursued as prospective viable and sustainable $H_2$ productions.[12-16] Recently, two dimensional (2D)-transition metal dichalcogenides (TMDs), whose crystal structure is composed of covalently bonded X-M-X (M = transition metal; X = S, Se, Te) layers which are held together by van der Waals forces,[17-23] have been reported as high-performance HER-electrocatalyst both in terms of electrocatalytic activity and stability.[24-34] Moreover, 2D-TMDs can be produced from their bulk crystal counterparts in suitable liquids to yield dispersions by liquid phase exfoliation (LPE).[35,36] This approach allows functional inks to formulate,[37] which can be processed by large-scale, cost-effective solution-based techniques,[38,39] offering the possibility to create and design layered artificial structures,[40,41] which have on-demand properties that are compatible with high-throughput industrial manufacturing.

Theoretical[42-45] and experimental[46-48] investigations have demonstrated that the unsaturated X-edges in the natural semiconducting 2H phase (trigonal prismatic) of TMDs are HER electrocatalytically active, and have a Gibbs free energy of adsorbed atomic H ($\Delta G_H^0$) close to zero.[42-44] These findings have rapidly promoted the development of nanostructured 2D-TMD electrodes with preferentially exposed edge sites,[34,42,43,47-50] which have shown state-of-the-art HER activity in the frame of noble metal-free electrocatalysts.[51]

Thermo-induced texturization processes in an $H_2$ environment have also been reported for activating the TMD basal planes by creating chalcogen-vacancies and forming edge-like sites,[52,53] *i.e.,* high HER-electrocatalytic activity was recorded across various morphologies, including bulk minerals, few-layer micro (lateral dimension of 2–5 μm) and nano (lateral dimension of ~200 nm) flakes.[53] Similarly, recent advances have demonstrated that the HER-electrocatalytic activity of 2D-TMDs can be significantly enhanced when their natural 2H phase is converted into a metallic 1T phase (distorted octahedral) via chemical exfoliation using organolithium compounds.[22,54-56] In fact, unlike the 2H phase, the basal plane of the 1T phase is also HER-electrocatalytically active.[57-59] However, the 1T phase is thermodynamically metastable, with a relaxation energy of ~1.0 eV[60] for the conversion to the stable 2H phase.[61,62] Moreover, the 2H-to-1T conversion method is complex and economically unviable.[63] In fact, it is typically achieved chemically by assisted Li-intercalation[18,58,64,65] or from multiple chemical reagents.[55,66,67]

In addition to the number of TMD HER-electrocatalytic active sites, their electron accessibility is also crucial to HER-performance.[34,68,69] Indeed, even though the electronic structure of the 2H-TMD edge is dominated by metallic one-dimensional states,[70] as owning density of states dominated by *d*-type orbitals,[71] hopping charge transport between the semiconducting states in the basal planes of 2H-TDMs[72] and the adjacent van der Waals bonded layers significantly reduces the overall HER-kinetics.[16,73-75] In order to overcome this limitation, the fabrication of TMD films with ultrathin (< 5 atomic layers) flakes[76,77] or with flakes that are stacked perpendicular to the conductive substrates[29,49] have been reported. Recently, the aforementioned electrical conductivity issue has been overcome by hybridizing 2D-TMDs with carbon-based conducting scaffolds, thus enabling faster HER-kinetics.[69,78-89] In particular, carbon nanotubes (CNTs)[89-95] and graphene/graphene derivatives[38,96-98] with high specific surface areas (up to 1315 $m^2$ $g^{-1}$[99,100] and 2630 $m^2$ $g^{-1}$[101-103] for single-wall CNTs (SWCNTs) and graphene, respectively), are ideal scaffold candidates.[104,105] Moreover, the design of graphene-/CNT-based hybrids effectively enhance the HER-electrocatalytic activity of the 2D-TMDs by reducing their $\Delta G_H^0$.[106] Lastly, the addition of CNTs and graphene to 2D-TMDs prevents agglomeration and restacking effects, determining a uniform distribution of 2D-TMDs within the carbon-based networks.[107-111] For example, $MoS_2$ flakes grown on reduced graphene oxide (RGO) flakes,[34] three dimensional (3D)-graphene network supporting perpendicularly-oriented $MoSe_2$,[112] 3D-$MoSe_2$ layered nanostructures grown on graphene flakes[84] and $MoSe_2$ flakes grown on the surfaces of porous N-doped CNT[87] all exhibited superior HER-electrocatalytic activity with respect to the individual 2D-TMDs. Recently, we reported that the HER-electrocatalytic activity of ~1 μm-thick $MoS_2$ flakes-based films (0.5 mg cm$^{-2}$ mass loading) on a self-standing graphene substrate is strongly enhanced with respect to the ones based on glassy carbon (GC) as a substrate.[113] Thus, a favourable electrochemical coupling of graphene with TMD flakes over a longer spatial range (*i.e.*, μm-scale) than that of hybrid composite materials (*i.e.,* nm-scale),[69,78-89] should also support the HER process. However, clear experimental and theoretical evidence of these long-range phenomena has not been reported yet.

Considering the aforementioned HER-findings for the 2D-TMDs and the possibility to formulate 2D material- and CNT-based inks,[37-39] we investigated solution-processed hybrid heterostructures made of either graphene flakes or SWCNTs and $MoSe_2$ flakes (henceforth named graphene/$MoSe_2$ and SWCNTs/$MoSe_2$, respectively) for HER (**Figure 1**). Among the TMDs, we opted for $MoSe_2$ because of its high intrinsic electrical conductivity (~$10^{-1}$ $\Omega^{-1}$ cm$^{-1}$)[114] with respect to that of the other TMDs (which is ~$10^{-2}$ $\Omega^{-1}$ cm$^{-1}$ for the most studied case of $MoS_2$[114]),[89] and low $\Delta G_H^0$ at its edges sites (<0.1 eV)[78,89,115]. As depicted in Figure 1, $MoSe_2$ and graphene flakes are produced in the form of dispersion by the LPE[35-39] of their bulk counterparts in 2-Propanol (IPA) and N-Methyl-2-pyrrolidone (NMP), respectively, which is followed by a sedimentation-based separation (SBS) process.[116-118] The SWCNT dispersions are produced by first dispersing SWCNTs in NMP then conducting an ultrasonication-based de-bundling process.[119-121] Subsequently, graphene/MoSe$_2$ or SWCNTs/MoSe$_2$ heterostructures are fabricated by sequentially depositing the as-formulated dispersions on nylon membranes through vacuum filtration. The optimization of the mass loading of the MoSe$_2$ flakes (up to 5 mg cm$^{-2}$) as well the electrode assembly *via* the monolithic stacking of different heterostructures provided remarkable HER-electrocatalytic activity (*i.e.*, overpotential at a cathodic current density of 10 mA cm$^{-2}$ ($\eta_{10}$) of 100 mV and a cathodic current density > 100 mA cm$^{-2}$ at an overpotential less than 200 mV).

Unlike HER-electrocatalysts based on 2D-TMDs/carbon-based material compounds,[78-89] the as-produced heterostructures have a μm-thick bilayer-like structure. Thus, the presented electrochemical results offer new insight into the HER-assisting coupling of 2D-TMDs with low-dimensional carbon materials over a longer spatial range than that of hybrid composite materials.[69,78-89] Moreover, electrode thermal annealing in an H$_2$ environment and chemical bathing in n-butyllithium are exploited to texturize the MoSe$_2$ flakes basal planes (through Se-vacancies creation), and to achieve *in situ* semiconducting-to-metallic phase conversion, respectively (*i.e.*, they activate new electrocatalytic sites). The as-engineered electrodes show faster HER-kinetics than those of untreated electrodes, which is evidenced by the Tafel plot analysis (~4.8-fold enhancement of the exchange current density ($j_0$) and a decrease in the Tafel slope from 63 to 54 mV dec$^{-1}$ after electrode chemical and thermal treatment, respectively).

To summarize, we provide methods and guidelines for producing and engineering TMD-based electrodes that are compatible with scalable manufacturing (*i.e.*, solution-based processing) and can compete with current noble metal-free technologies, marking a turning point in electrochemical H$_2$ production through 2D-TMD-based electrocatalysts.

## 2. Results and Discussion
### 2.1. Production, Processing and Characterization of SWCNTs, Graphene and MoSe$_2$

The MoSe$_2$ flakes are produced by the LPE[35,37] of the MoSe$_2$ bulk crystal in IPA (see Experimental Section), thus avoiding typical problems relating to the LPE of TMDs in toxic and high boiling point solvents such as NMP and dimethylformamide (DMF), including high-temperature annealing processing for solvent removal as well as surface oxidation.[122,123,124] The morphology of the MoSe$_2$ flakes, as selected by the SBS process[116-118] (see Experimental Section) is characterized by transmission electron microscopy (TEM) and atomic force microscopy (AFM) (**Figure 2**) in order to evaluate their lateral dimension and thickness, respectively. Figure 2a shows the TEM image of the MoSe$_2$ flakes, displaying a crumpled paper-like structure. Statistical TEM analysis (Figure 2b) indicates a lateral size of the flakes in the range of 10-170 nm (log normal distribution peaks at ~29 nm). An additional TEM image of a single MoSe$_2$ flake is reported in the Supporting Information (S.I.) (**Figure S1**a), together with the TEM-selected area electron diffraction (TEM-SAED) (Figure S1b). The latter shows a sharp ring-and-dot pattern, indicating the polycrystalline-like nature of the MoSe$_2$ flakes. A representative AFM image of the exfoliated MoSe$_2$ flakes is shown in Figure 2c, together with the height profile of a single MoSe$_2$ flake (white line in Figure 2c), showing nano-edge steps (*i.e.,* flake thickness) of ~1 nm. Additional AFM images are reported in the S.I. (Figure S1c,d). Statistical AFM analysis (Figure 2d) evidences the presence of single- to a few-layer MoSe$_2$ flakes (the thickness of a MoSe$_2$ monolayer lies generally between 0.6 nm and 1 nm[Q25,126]), with a log normal distribution peaking at ~3 nm.

Optical absorption spectroscopy (OAS) of MoSe$_2$ flake dispersion in IPA is reported in **Figure S2**. Absorption peaks around 810 nm (1.53 eV) and 708 nm (1.75 eV) correspond to the A and B excitonic peaks. These peaks arise from direct inter-band transitions at the K-point in the Brillouin zone of the 2H-phase MoSe$_2$.[127-130] the latter originate from the energy split of the valence-band that is formed from the Mo atom[131,132] and spin-orbital coupling due to the in-plane confinement of the electron and the atomic mass of Mo.[132,133] The shoulder in the absorption spectra around ~410 nm is attributed to the C and D inter-band transitions between the density of state peaks in the valence and conduction bands of the 2H-phase of MoSe$_2$.[134,135]

Raman spectroscopy is carried out on both MoSe$_2$ bulk and MoSe$_2$ flakes in order to investigate their different topological structures.[136] According to group theory analysis, bulk TMDs are members of D$_{6h}$ point group symmetry,[137] and are characterized by four Raman active modes, *i.e.,* three in-plane E$_{1g}$, E$^1_{2g}$, and E$^2_{2g}$, and one out-of-plane A$_{1g}$. Only two of these are typically accessible experimentally, namely E$^1_{2g}$ and A$_{1g}$[138] since the E$^2_{2g}$ mode is at very low frequency (~30 cm$^{-1}$), and the E$_{1g}$ mode is forbidden in backscattering geometry on a basal plane.[138] Additionally, when the number of layers decreases below a certain threshold, the interlayer vibrational mode B$_{2g}$ becomes active as a result of the breakdown of the translational symmetry.[139-140] This mode is present only in few-layered flakes, not in single-layered MoSe$_2$.[141] Representative spectra of MoSe$_2$ bulk and MoSe$_2$ flakes are reported in **Figure 3**a. The A$_{1g}$ mode is located at ~241 cm$^{-1}$ for the MoSe$_2$ bulk, while it is red-shifted to ~239 cm$^{-1}$ for the MoSe$_2$ flakes, which is in agreement with the softening of the vibrational mode that is accompanied by the flake thickness reduction.[141-146] The in-plane E$^1_{2g}$ mode is observed at ~287 cm$^{-1}$ for both samples.[141-144] In the case of MoSe$_2$ flakes, the E$^1_{2g}$ peak position (Pos(E$^1_{2g}$)) and intensity (I(E$^1_{2g}$)) are estimated by simultaneously fitting the E$^1_{2g}$ and the close (partially overlapping) Si peaks.[147] This procedure is not applied for MoSe$_2$ bulk because its I(E$^1_{2g}$) is lower than that of the MoSe$_2$ flakes. The intensity ratio between the A$_{1g}$ and E$^1_{2g}$ modes (I(A$_{1g}$)/I(E$^1_{2g}$)) for MoSe$_2$ flakes is ~21. This value is consistent with those reported for few-layered MoSe$_2$ flakes.[143,148] The in-plane E$_{1g}$ mode is observed at ~167 cm$^{-1}$ in both MoSe$_2$ bulk and MoSe$_2$ flakes. The activation of the E$_{1g}$ mode is linked with a resonance-induced symmetry breaking effect.[149] Moreover, the energy of this mode, which is independent of the number of layers,[150] does not change between the MoSe$_2$ bulk and exfoliated flakes. Finally, the B$_{2g}$ mode, which is inactive for MoSe$_2$ bulk, is present at

~352 cm$^{-1}$ for the MoSe$_2$ flakes,[141] confirming their few-layered composition. The statistical Raman analysis is reported in the S.I. (**Figure S3**).

The crystallinity of the MoSe$_2$ flakes is investigated through X-ray powder diffraction (XRD). Figure 3b show the XRD pattern obtained for the MoSe$_2$ flakes, together with that of the MoSe$_2$ bulk. The latter can be indexed with the JCPDS Card No. 29-0914 of the hexagonal phase of MoSe$_2$ (*i.e.,* 2H-MoSe$_2$), which is in agreement with several reports in literature.[151-153] In the case of MoSe$_2$ flakes, the (002) peak is clearly broader (see inset to Figure 3b) and the other peaks, although retaining the same position of the native bulk, are strongly reduced in intensity. This indicates the exfoliation along the c-axis of the MoSe$_2$ flakes, which occurs without any phase changes.[113,154,155] The data also exclude the presence of Se powder and oxidized species crystal domains, since their corresponding XRD peaks are not present.

X-ray photoelectron spectroscopy (XPS) measurements are performed on both MoSe$_2$ bulk and MoSe$_2$ flakes to further study the chemical material composition and oxidation states of the elements. Mo 3d and Se 3d XPS spectra are shown in Figures 3c and 3d, respectively, together with their deconvolution. In Figure 3c, the two peaks located at (229.3±0.2) eV and (232.4±0.2) eV correspond to the Mo 3d$_{5/2}$ and Mo 3d$_{3/2}$ peaks of the Mo(IV) state in MoSe$_2$, which is in agreement with literature reports on MoS$_2$ and MoSe$_2$.[156-160] The additional peaks at binding energies of 232.5±0.2 eV and 235.7±0.2 eV are assigned to the Mo(VI) state and are related to MoO$_3$ residue.[161-163] The compositional analysis indicates that the percentage content (%c) of MoO$_3$ (defined as MoO$_3$/(MoO$_3$+MoSe$_2$)) is ~11% in the case of the MoSe$_2$ flakes, which is consistent with the one recorded for MoSe$_2$ bulk (~6%). This result proves that the LPE of MoSe$_2$ bulk crystal in IPA produces MoSe$_2$ flakes, but since it does not have the same drawbacks as the TMD flakes produced by LPE in NMP,[164,165] oxidized species at significantly higher contents are generated (%c is 40-60%,[164,165] depending on the exfoliation process[165]). In Figure 3d, the peaks at 54.9±0.2 eV and 55.7±0.2 eV are attributed to the Se 3d$_{5/2}$ and Se 3d$_{3/2}$ peaks, of the diselenide moiety of MoSe$_2,$ respectively.[158-160,166-168] The compositional analysis shows that the Mo:Se atomic ratio is ~1:1.9 for both $MoSe_2$ bulk and $MoSe_2$ flakes, which is almost in agreement with the theoretical stoichiometry of $MoSe_2$ (1:2). The graphene flake dispersions are produced by LPE,[35,37] followed by the SBS[116-118] of pristine graphite in NMP, while SWCNT dispersions are produced by dispersing SWCNTs in NMP with the aid of ultrasonication for the de-bundling process.[119,120] Additional details concerning the dispersion production are reported in the Experimental Section, following procedures previously reported.[113,169,170]

The morphological, optical and structural properties of the as-produced graphene flakes and SWCNTs are reported in the S.I. (**Figure S4-S8**), and are consistent with results shown in our previous works.[113,169,170]

## 2.1. Solution-processed Graphene/$MoSe_2$ and SWCNT/$MoSe_2$ Heterostructures

In order to take advantage of the favourable physical and electrochemical coupling of graphene or SWCNTs with $MoSe_2$ flakes for enhancing HER-electrocatalytic activity, solution-processed hybrid heterostructures (*i.e.,* graphene/$MoSe_2$ and SWCNTs/$MoSe_2$) are produced by sequentially depositing the as-produced material dispersions on nylon membranes through vacuum filtration. A mass loading of 2 mg cm$^{-2}$ is first adopted for all three materials (graphene flakes, SWCNTs and $MoSe_2$ flakes). Afterward, the mass loading of the $MoSe_2$ flakes is increased to 5 mg cm$^{-2}$ in order to increase the overall number of active sites through controlling the electrode thickness, as it has been recently observed for liquid phase exfoliated TMD-based electrodes.[88,89] Furthermore, electrodes made of graphene flakes or SWCNTs only (henceforth named graphene and SWCNTs, respectively), *i.e.,* electrodes without the $MoSe_2$ flake deposition, are also produced as points of reference. The details of the fabrication of the electrodes are reported in the Experimental Section.

The surface morphology of the as-prepared electrodes is characterized by scanning electron microscopy (SEM) and AFM. **Figures 4**a-d show the top-view SEM images of the electrodes.

The surface of the graphene electrode has a crumpled, wrinkled and flake-like structure, while the surface of the SWCNT electrode consists of a mesoporous network forming a bundle-like morphology. The surface of the hybrid electrodes (MoSe$_2$ flakes mass loading = 2 mg cm$^{-2}$) is clearly modified by the addition of MoSe$_2$ flakes with respect to the bare graphene and SWCNT electrodes. For the graphene/MoSe$_2$ electrode, the MoSe$_2$ flakes uniformly cover the underlying graphene flakes. In contrast, the underlying mesoporous network of the SWCNTs is still observed for SWCNTs/MoSe$_2$, demonstrating that the MoSe$_2$ flakes are penetrated between the SWCNTs.

**Figure S9** reports the AFM images of the electrode surfaces, evidencing morphologies that are similar to those observed by SEM. The average roughness (Ra) values are ~46.2 nm for graphene electrodes and ~103 nm SWCNT electrodes. These values decrease to ~21 nm and 70 nm for the corresponding hybrid electrodes, respectively, indicating that MoSe$_2$ flake deposition flattens the electrode surfaces. For graphene/MoSe$_2$, the flattening of the electrode surface is attributed to the lateral dimension of the MoSe$_2$ flakes (10-170 nm) which is smaller than that of graphene flakes (200-1500 nm). This leads to a more compact (*i.e.*, more dense) overlayer with respect to the film base graphene flakes. For the SWCNTs/MoSe$_2$, the surface flattening is a consequence of the MoSe$_2$ flake filling and the coverage of the SWCNT mesoporous network.

Figures 4e-h show the high-resolution cross-sectional SEM images of the representative graphene/MoSe$_2$ and SWCNTs/MoSe$_2$. For the graphene/MoSe$_2$, a well-defined bilayer structure is observed (Figure 4e,f). MoSe$_2$ flakes are deposited as a homogeneous porous overlayer (see the high-magnification image, Figure 4f) because of the filter-like behaviour of the graphene flakes.[171-173] The estimated layer thickness is ~2.5 μm and ~0.8 μm for the graphene and MoSe$_2$ flake layer, respectively. Considering that the graphene and MoSe$_2$ flakes have an identical mass loading, these values indicate that the film of the MoSe$_2$ flakes is denser than that of the graphene flakes. For SWCNTs/MoSe$_2$, the high-magnification SEM

image (Figure 4g) reveals that the MoSe$_2$ flakes penetrate the mesoporous SWCNT network (commonly referred to as "buckypaper").[174,175] However, a bilayer structure is still observed, as is also confirmed by elemental energy-dispersive X-ray spectroscopy (EDX) analysis of the cross-sectional SEM images of SWCNTs/MoSe$_2$ (**Figure S10**). Moreover, the thickness of the whole electrode exceeds 100 μm, which indicates a low-density "buckypaper" formation. **Figure S11** shows the Raman spectra of the graphene/MoSe$_2$ and SWCNTs/MoSe$_2$, focusing on the 140-410 cm$^{-1}$ spectral region, where the MoSe$_2$ flakes Raman peaks are located (see Figure 3a). The comparison with the Raman spectrum of the MoSe$_2$ flakes does not reveal any significant differences, which suggests that no structural modifications of the as-produced-MoSe$_2$ flakes occur during film deposition through the vacuum filtration of their dispersions. Similar conclusions are also derived from XPS measurements on the heterostructures (**Figure S12**), since no significant differences are observed in the Mo 3d and Se 3d spectra with respect to those of the as-produced MoSe$_2$ flakes. In particular, XPS analysis reveals that no relevant MoSe$_2$ flake oxidation occurred (%c of MoO$_3$: ~12% for graphene/MoSe$_2$; ~17% for SWCNTs/MoSe$_2$) and that there is a slight binding energy downshift of the MoSe$_2$-related peaks (~0.1 eV for graphene/MoSe$_2$ and ~0.2 eV for SWCNTs/MoSe2) with respect to those of the as-produced MoSe$_2$ flakes.

**2.3. Electrochemical characterization**

The HER-electrocatalytic activity of the as-produced electrodes is evaluated in 0.5 M H$_2$SO$_4$ (pH 1), in which the MoSe$_2$ flakes show higher HER-electrocatalytic activity than that achieved under alkaline conditions (*e.g.*, 1 M KOH, pH 14, see **Figure S13**). Actually, the kinetic energy barrier of the initial Volmer step and the strong adsorption of the formed OH$^-$ on the surfaces of MoSe$_2$ flakes (and, more generally, of the 2D-TMDs) are considered responsible for the sluggish HER kinetics in alkaline solutions (see additional details in the S.I.).[176,177] The MoSe$_2$ flakes are also deposited and tested on a GC electrode (*i.e.,*

GC/MoSe$_2$) in order to test their native electrocatalytic properties on a flat, inert, conductive substrate.

**Figure 5**a displays the *iR*-corrected polarization curves for the different electrodes. Clearly, the graphene/MoSe$_2$ and SWCNTs/MoSe$_2$ have a higher current density than the GC/MoSe$_2$. The overpotential *vs.* reversible hydrogen electrode (RHE) at a 10 mA cm$^{-2}$-cathodic current density ($\eta_{10}$) decreases from 0.34 V for GC/MoSe$_2$ to 0.18 V for graphene/MoSe$_2$ and 0.17 V for SWCNTs/MoSe$_2$. Notably, the appreciable cathodic current density at 0 V *vs.* RHE for the graphene, SWCNTs and the heterostructures is the capacitive current (for 5 mV s$^{-1}$ potential scan rate, see Methods section) attributed to the high gravimetric capacitance of the carbon nanomaterials and TMDs coupled with graphene or SWCNTs (hundreds of F g$^{-1}$).

The Tafel slope and j$_0$ are also useful Figures of Merit (FoM) to assess the HER-electrocatalytic activity.[178,179] These FoM are extracted from the linear portion of the Tafel plot, which shows the relation between the overpotential and the current density of the electrodes, which is consistent with the Tafel equation (see Experimental Section).[178,179] The Tafel slope measures the potential increase that is required to improve the current density by 1 order of magnitude.[178,179] Fundamentally, it is used to determine the HER mechanism at the electrode/electrolyte interface.[178,179] Possible HER-pathways involve the Volmer reaction (H$_3$O$^+$ + *e*$^-$ → H$_{ads}$ + H$_2$O, in which H$_{ads}$ refer to adsorbed atomic hydrogen on the electrocatalyst) and either the Heyrovsky reaction (H$_{ads}$ + H$_3$O$^+$ + *e*$^-$ → H$_2$ + H$_2$O) or Tafel reaction (H$_{ads}$ + H$_{ads}$ + *e*$^-$ → H$_2$), to give either a Volmer-Heyrovsky[178,179] or a Volmer-Tafel mechanism.[178,179] In the case of an insufficient H$_{ads}$ surface coverage, the Volmer reaction is the rate limiting step of the HER and a theoretical Tafel slope of 120 mV dec$^{-1}$ is expected.[178,179] On the contrary, in the case of a high H$_{ads}$ surface coverage, the HER-kinetic is dominated by the Heyrovsky or Tafel reaction, and a Tafel slope of 40 or 30 mV dec$^{-1}$ is detected.[178,179] The j$_0$ is positively correlated to the number of available HER-electrocatalytic sites and their HER-kinetics.[178-179]

The Tafel slopes are 88, 80, and 67 mV dec$^{-1}$ for GC/MoSe$_2$, graphene/MoSe$_2$ and SWCNTs/MoSe$_2$, respectively. These values are all in agreement with the Volmer-Heyrovsky HER-mechanism, which is consistent with previous studies on 2D-TMDs.[24-34] However, a decrease in the Tafel slope value is observed for graphene/MoSe$_2$ and SWCNTs/MoSe$_2$ with respect to that observed for GC/MoSe$_2$, indicating that the HER-electrocalytic activity of the electrode is barely limited by the Volmer reaction step.

The j$_0$ values are 5, 56 and 29 µA cm$^{-2}$ for GC/MoSe$_2$, graphene/MoSe$_2$ and SWCNTs/MoSe$_2$. The value obtained for GC/MoSe$_2$ is consistent with those reported in literature for 2D-TMDs with similar mass loadings.[88,89] The values obtained for SWCNTs/MoSe$_2$ are comparable to those reported for MoSe$_2$ flake/SWCNT compounds (in the order of 10$^2$ µA cm$^{-2}$).[89] Notably, the highest j$_0$ value is measured for graphene/MoSe$_2$, which is a bilayer-like heterostructure consisting of a graphene flake film covered by an homogeneous layer of MoSe$_2$ flakes (see SEM analysis, Figures 4e,f).

The aforementioned results indicate two HER-assisting properties of the heterostructures: 1) the electrical conductivity of the MoSe$_2$ flakes guarantees the electron accessibility to HER-electrocatalytic sites in a film with a µm-thickness scale; 2) the overall kinetics of the MoSe$_2$ flakes are accelerated, with respect to the GC/MoSe$_2$, by the favourable electrochemical coupling with graphene flakes or SWCNTs substrates which cause a decrease in MoSe$_2$ flakes ΔG$_H^0$. Notably, this electrochemical coupling is effective for ~µm-thick layers of MoSe$_2$ flakes, which is different to the short-spatial range coupling expressed by hybrid graphene/ or CNTs/2D-TMDs.[78-89,106] The binding energy downshift of the MoSe$_2$-related XPS peaks (see Figure S12) might be ascribed to the influence of the graphene or SWCNTs on the electronic state of the deposited MoSe$_2$ flakes, which weakens the ΔG$_H^0$ on the MoSe$_2$ flakes, increasing the HER electrocatalytic performance. This is in agreement with a recently reported simulated deformation charge density of the MoS$_2$/SWCNT interface, which shows that 0.924 electrons can be transferred from SWNTs to $MoS_2$.[69] In addition, for the $SWCNTs/MoSe_2$, the porosity of the $MoSe_2$ flake overlay, as observed by SEM analysis (Figure 4g-h), also supports the $H_{ads}$ surface coverage, thus the $\Delta G_H^0$ (*i.e.*, the Tafel slope) can be reduced further compared to graphene/$MoSe_2$. Lastly, the interpenetration of the SWCNTs and $MoSe_2$ flakes for the $SWCNTs/MoSe_2$ (see SEM analysis, Figures 4g,h) is expected to increase the electron accessibility of the HER-electrocatalytic sites, with the electrode conductivity being enhanced by the presence of SWCNTs.[78,88,89]

Considering the aforementioned experimental results, the increase in the mass loading of $MoSe_2$ flakes (up to 5 mg cm$^{-2}$) (**Figure 6**a, left sketch) in $SWCNTs/MoSe_2$ is driven by the need to fully exploit the HER-assisting properties of the heterostructures. For mass loading exceeding 5 mg cm$^{-2}$, material films resulted unstable against fragmentation during preparation. However, the monolithic stacking of different heterostructures ($SWCNTs/MoSe_2$) (Figure 6a, right sketch) is proposed as a smart electrode assembly for achieving state-of-the art HER-performance (*e.g.*, a cathodic current density > 100 mA cm$^{-2}$ at an overpotential less than 0.2 V),[180,181] thus overcoming the mass loading-related limit of the HER-performance in the single heterostructure.

As reported in Figures 6b-c, the as-produced electrodes show remarkable HER-electrocatalytic activity, *i.e.*, $\eta_{10}$ values of 0.15, 0.12 and 0.10 V for 1, 2 and 6 monolithically stacked $SWCNTs/MoSe_2$, each one with a mass loading of 5 mg cm$^{-2}$ for $MoSe_2$ flakes (*i.e.*, a total electrode mass loading of 5, 10, and 30 mg cm$^{-2}$, respectively). The $\eta_{10}$ reduction is explained by the Tafel plot analysis, which marks an increment of $j_0$ with an increase in both the mass loading of the $MoSe_2$ flakes (from 2 to 5 mg cm$^{-2}$) and the number of stacked electrodes (from 1 to 6), meaning an effective increase in the HER-electrocatalytic active sites of the $MoSe_2$ flakes. The $j_0$ values are 64, 165 and 203 μA cm$^{-2}$ for the electrode with a mass loading of 5 mg cm$^{-2}$, and for those obtained by stacking 2 and 6 electrodes, respectively. It is worth noting that the $j_0$ value of 203 μA cm$^{-2}$ is one of the highest values reported in literature for TMDs,[88,89] exceeding even those that are usually reported for 1T-TMDs (*e.g.,* 167 µA cm$^{-2}$ for 1T-MoS$_2$ nanoparticles[182]).[47,183] Tafel slope values, however, are similar for all the electrodes, thus suggesting that the same HER-mechanism (Volmer-Heyrovsky) occurs.

## 2.4. Engineering of the HER-electrocatalytic MoSe$_2$ flakes

In order to improve the HER-electrocatalytic activity of the MoSe$_2$ flakes, and of 2D-TMDs in general, it is essential to activate their basal plane,[46-50] *i.e.,* the inert (0001) surface plane,[42-48] as well to increase their electrical conductivity.[72-75] Thus, two treatments are investigated (**Figure 7**): 1) thermo-induced texturization by annealing flakes in an H$_2$ environment; 2) *in situ* semiconducting (2H-MoSe$_2$)-to-metal (1T-MoSe$_2$, MoO$_x$ and elemental atoms) phase conversion by chemical bathing flakes in an organo-lithium compound.

In the first process, Se atoms in defect-free MoSe$_2$ flakes (Figure 7a) are expected to be removed as H$_2$Se gas,[184,185] leading to the formation of Se-vacancies and edges in the (0001) plane.[52,53] Simultaneously, the excess Mo could form metal clusters on the MoSe$_2$ flakes (Figure 7b). A similar treatment has been reported for MoS$_2$, in which the HER-electrocatalytic activity was improved by increasing the edge site intensity, whose *d*-type orbitals are known to actively participate in catalysed reaction,[71,186] on the surface and by making the flake conductive through the generation of metallic Mo clusters.[52,53]

The second treatment is expected to induce the 2H-to-metallic (1T-MoSe$_2$, MoO$_x$ and elemental atoms) phase conversion of the MoSe$_2$ (Figure 7c). A similar approach has been applied to MoS$_2$ monolayer-based field-effect transistors (FETs) in order to locally induce the 2H-to-1T phase conversion of the MoS$_2$.[187] This phase engineering decreased the high-resistance contacts (0.7 kΩ µm-10 kΩ µm) of 2H-MoS$_2$ to 200-300 Ω µm, thus it optimized the injection of the charge carriers into the channel.[187] In our case, the phase conversion of MoSe$_2$ flakes is expected to increase their electron conductivity and enhance the surface reactivity of their basal plane for atomic H binding (*i.e.,* to decrease $\Delta G_H^0$). This, in turn, might facilitate the Volmer reaction step, which favors the subsequent Heyrovsky reaction.[54-57]

The first treatment is preliminary applied to the GC/MoSe$_2$ in order to investigate the effect of different annealing temperatures on the HER-electrocatalytic activity of MoSe$_2$ flake films. X-ray photoelectron spectroscopy measurements (**Figure S14**) show the progressive formation of elemental Mo(0) following the increase in the annealing temperature. In more detail, the %c of Mo(0) and the total Se is > 10% and < 20%, respectively, for an annealing temperature ≥ 700 °C. Under these conditions, Mo (VI) is also observed with %c > 50% which might be attributed to the subsequent oxidation of the elemental Mo under air exposure.[188] X-ray photoelectron spectroscopy analysis is carried out on MoSe$_2$ flake films for evaluating the effects of the chemical treatment, that is 12 h-chemical bathing in n-butyllithium (**Figure S15**). The results confirm the modification of the surface chemistry of MoSe$_2$. The spectra show the formation of different metallic phases (*e.g.,* MoO$_x$ and Mo) and additional elemental atoms (Se and residual Li), which overlap and contribute to the Mo 3d and Se 3d spectra of MoSe$_2$ flakes (Li-species 1s XPS spectrum peaks between 50-60 eV), respectively. MoSe$_2$-related XPS bands are attributed to both the semiconducting (2H) and metallic (1T) phases.

**Figure 8** reports the AFM images of the annealed MoSe$_2$ flake films deposited on Si substrate, in comparison with the not annealed ones. The results show a progressive size reduction in the MoSe$_2$ flakes, and a consequent smoothing of their films, when the annealing temperature is increased up to 700 °C. In fact, the Ra value reduces from 22 nm for the film that was not annealed, to 12 and 11 nm for the films that were annealed at 600 °C and 700 °C, respectively. However, when the temperature is further increased to 800 °C, the aggregates form due to the excessive removal of Se and Mo cluster assemblies, which consequently determines a Ra increase up to 20 nm.

**Figure S16** reports the electrochemical characterization of the electrodes annealed at 600, 700 and 800 °C in Ar/$H_2$ (90/10%) for 5 h, and compares it with that of the untreated electrode. The results evidence that the HER-electrocatalytic activity of the electrodes annealed at 600 and 700 °C is enhanced with respect to the untreated electrode. In particular, $\eta_{10}$ decreases from 0.34 V in the untreated electrode to 0.29 and 0.26 in the electrodes annealed at 600 and 700 °C, respectively. A further increase in the temperature up to 800 °C causes a deterioration of the HER-electrocatalytic activity, whose $\eta_{10}$ (0.44 V) increases by 0.1 V compared to that of the untreated electrode. Tafel slope values are also positively affected by the thermal treatment at 600 and 700 °C, for which they are 86 and 74 mV dec$^{-1}$, respectively. The lowest Tafel slope is observed (~144 mV dec$^{-1}$) for the treatment at 800 °C. The $j_0$ values increase for all the tested annealing temperatures, (19, 11 and 10 μA cm$^{-2}$ for 600, 700 and 800 °C). These results confirm the effectiveness of the texturization of the basal plane of $MoSe_2$ flake films due to the formation of Se-vacancies (*i.e.,* HER-electrocatalytic sites) that are caused by $H_2Se$ gas evolution, as stated above. This is also confirmed by Electrochemical Impedance Spectroscopy (EIS) analysis, which correlates the increase double-layer capacitance (Cdl) of the electrodes with the texturization process after their thermal annealing (see S.I., **Figure S17**), i.e., with their porosity.[189] However, at the highest annealing temperature of 800 °C, the excessive removal of Se deteriorates the $MoSe_2$ phase, thus the HER-electrocatalytic activity decreases.

Driven by the obtained results on GC/$MoSe_2$, we subsequently treated the SWCNTs/$MoSe_2$-based electrode, which has shown the best HER-electrocatalytic activity (see Figure 5). We carried out annealing in $H_2$ at 700 °C for a 12 h-chemical bathing in n-butyllithium.

**Figure 9**a displays the polarization curves obtained for the treated electrodes and compares them with that of the untreated one ($MoSe_2$ flake mass loading of 2 mg cm$^{-2}$). Clearly, both chemical and thermal treatments enhance the HER-electrocatalytic activity of the electrode. In detail, the $\eta_{10}$ decreases from 0.17 V for the untreated electrode to 0.15 and 0.13 V for the Li-

intercalated electrodes and the electrodes annealed in H$_2$, respectively. The Tafel slope and j$_0$ values are 83 mV dec$^{-1}$, and 167 μA cm$^{-2}$, respectively, for the chemically treated electrode, and 54 mV dec$^{-1}$ and 55 μA cm$^{-2}$, respectively, for the thermally treated one. Notably, the treated electrodes show a remarkable increase in j$_0$ (479% and 90% after chemical and thermal treatments, respectively) with respect to that of the untreated electrode (j$_0$ = 29 μA cm$^{-2}$). This indicates an increase in the number of the HER-electrocatalytic sites as a result of the semiconducting-to-metallic phase conversion of the MoSe$_2$ flakes and the addition of Se-vacancies to their basal planes after chemical and thermal treatments, respectively.[52,53] Moreover, the thermal treatment decreases the Tafel slope values (from 67 mV dec$^{-1}$ to 54 mV dec$^{-1}$), therefore it is an effective method for enhancing the overall HER-kinetics. The electrochemical stability of the untreated graphene/MoSe$_2$, SWCNTs/MoSe$_2$ and the treated electrodes is evaluated by chronoamperometry measurements (*j-t* curves). In all cases, a constant overpotential is applied in order to give an equal starting current density of -30 mA cm$^{-2}$, which is similar to operative HER-conditions. As shown in **Figure S18**, the electrodes retain a steady HER-electrocatalytic activity over a period of 40 000 s (*i.e.,* > 11 h). In particular, for the SWCNTs/MoSe$_2$ chemically treated in n-butyllithium, the current density decreases by ~28%. The HER-electrocatalytic activity degradation might be caused by the thermodynamically metastable nature of the 1T-phase, which could be converted back to the native 2H-phase,[60-62] or by the dissolution of soluble MoO$_x$ species in acid.[190-191] However, for the untreated graphene/MoSe$_2$, the SWCNTs/MoSe$_2$ and the SWCNTs/MoSe$_2$ annealed at 700 °C in an H$_2$ environment, only slight current density fluctuations are observed, which might as a reault of the consumption of H$^+$ or the accumulation of H$_2$ bubbles on the electrode surface, which hinders the HER.[47,76,145] Thus, these electrodes fulfill long-durability requirement, as expected from the 2H-phase of MoSe$_2$.[60-62] In order to provide additional insight on the electrochemical stability of the MoSe$_2$, XPS measurements are carried out on the heterostructures, as well on the GC/MoSe$_2$, after the electrochemical stability test. **Figure S19** reports the XPS measurements of the surfaces of the GC/MoSe, graphene/MoSe$_2$ and SWCNT/MoSe$_2$ before and after the electrochemical stability tests. These data confirm that no significant changes occur for the MoSe$_2$-related bands both in the Mo 3d and Se 3d XPS spectra, thus confirming that the 2H-phase of the MoSe$_2$ flakes is electrochemically stable during HER-operation. Furthermore, after HER, the MoO$_x$ species are still not observed, suggesting their dissolution in acid.[190,191] This effect could be effective for improving the exposure of the MoSe$_2$ active sites, thus favoring the HER-activity of the electrodes.

## 3. Conclusions

Solution-processed heterostructures based on MoSe$_2$ flakes and either graphene flakes or single-wall carbon nanotubes (SWCNTs) are produced for an efficient HER. All the nanomaterials are produced and processed in the form of liquid dispersions, which are compatible with the fabrication of high-throughput scalable electrodes. The heterostructure-based approach for designing HER-electrocatalysts permits the optimization of the exfoliating solvent and protocols for each nanomaterial independently of each other. The μm-spatial range electrochemical coupling of the MoSe$_2$ flakes with graphene flakes or SWCNTs increases the HER-electrocatalytic activity of the MoSe$_2$ flakes. In particular, a remarkable $\eta_{10}$ of 100 mV and a cathodic current density > 100 mA cm$^{-2}$ at an overpotential less than 200 mV are achieved by optimizing the mass loading of MoSe$_2$ flakes on SWCNTs and by electrode assembly *via* the monolithic stacking of multiple heterostructures. Our proposed approach permits to obtain self-standing electrodes, which does not require additional active film transfer process on conductive current collectors. By selecting the number of the heterostructures, the mass loading of the active materials is no longer limited as in the single heterostructure. In the latter case, material mass loading exceeding 5 mg cm$^{-2}$ determines active films fragmentation, occurring during deposition and/or HER-operation. Moreover, electrode thermal annealing in an $H_2$ environment is conducted for texturizing the basal plane of the $MoSe_2$ flakes, while electrode chemical bathing in n-butyllithium effectively triggers the *in situ* semiconducting-to-metallic phase conversion of the $MoSe_2$ flakes. Both treatments create new HER-electrocatalytic sites in the $MoSe_2$ flakes. Consequently, the SWCNTs/$MoSe_2$ show a ~4.8-fold enhancement of the $j_0$ (from 29 to 167 $\mu A\ cm^{-2}$) after chemical treatments, and a ~20% decrease in the Tafel slope (from 67 to 54 $mV\ dec^{-1}$) after thermal annealing at 700 °C in Ar/$H_2$ (90/10%), respectively. The untreated and thermally treated heterostructures fully retain steady HER-electrocatalytic activities for more than 11 h, thus they meet the durability requirements for practical applications.

The proposed engineering strategies can be generally extended to other liquid-exfoliated HER-electrocatalytic 2D materials, thus providing general guidelines to design state-of-the-art electrodes for large-scale electrochemical $H_2$ production.

**4. Experimental Section**

*Production and processing of materials*

The $MoSe_2$ flakes are produced by LPE,[35,37] followed by SBS,[116-118] in 2-Propanol (IPA) of $MoSe_2$ bulk crystal. In short, 30 mg of $MoSe_2$ bulk crystals are added to 50 mL of IPA and ultrasonicated in a bath sonicator (Branson® 5800 cleaner, Branson Ultrasonics) for 6 h. The resulting dispersion is ultracentrifuged at 2700 g (in a Beckman Coulter Optima™ XE-90 with a SW32Ti rotor) for 60 min at 15 °C in order to separate un-exfoliated and thick $MoSe_2$ crystals (collected as sediment) from the thin $MoSe_2$ flakes that remain in the supernatant. Then, 80% of the supernatant is collected by pipetting, obtaining dispersion of $MoSe_2$ flakes. The graphene flakes are produced by the LPE,[35,37] followed by SBS,[116-118] of pristine graphite (+100 mesh, ≥75% min, Sigma Aldrich) in NMP. Experimentally, 1 g of graphite is dispersed in 100 ml of NMP (99.5% purity, Sigma Aldrich) and ultrasonicated in a bath sonicator (Branson® 5800 cleaner, Branson Ultrasonics) for 6 h. The resulting dispersion is then ultracentrifuged at 17000 g (in Beckman Coulter Optima™ XE-90 with a SW32Ti rotor) for 50 min at 15 °C to exploit SBS. Next, ~80% of the supernatant is collected by pipetting, obtaining dispersion of graphene flakes.

The SWCNTs (> 90% purity, Cheap Tubes) are used as received, without any purification steps. The SWCNT dispersions are produced by dispersing SWCNTs in NMP at a concentration of 0.2 g L$^{-1}$ using ultrasonication-based de-bundling.[119,120] In short, 10 mg of SWNT powder is added to 50 mL of IPA in a 100 mL open topped, flat-bottomed beaker. The dispersion is sonicated using a horn probe sonic tip (Vibra-cell 75185, Sonics) with the vibration amplitude set to 45% and a sonication time of 30 min. The sonic tip is pulsed at a rate of 5 s on and 2 s off to avoid damage to the processor and to reduce any solvent heating. An ice bath around the beaker is used during sonication in order to minimize heating effects.

*Material characterization*

Transmission electron microscopy images are taken with a JEM 1011 (JEOL) transmission electron microscope operating at 100 kV. Samples for the TEM measurements are prepared by drop-casting the MoSe$_2$ flakes, graphene flakes and SWCNT dispersions onto carbon-coated Cu grids, before rinsing them with deionized water and subsequently drying them under vacuum overnight. Morphological and statistical analysis is carried out by using ImageJ software (NIH) and OriginPro 9.1 software (OriginLab), respectively.

Atomic force microscopy images are taken using a Nanowizard III (JPK Instruments, Germany) mounted on an Axio Observer D1 (Carl Zeiss, Germany) inverted optical microscope. The AFM measurements are carried out by using PPP-NCHR cantilevers (Nanosensors, USA) with a nominal tip diameter of 10 nm. A drive frequency of ~295 kHz is used. Intermittent contact mode AFM images (512x512 data points) of 2.5×2.5 µm$^2$ and 500×500 nm$^2$ are collected by keeping the working set point of the free oscillation amplitude above 70%. The scan rate for the acquisition of images is 0.7 Hz. Height profiles are processed by using JPK Data Processing software (JPK Instruments, Germany) and the data are analysed with OriginPro 9.1 software. Statistical analysis is carried out by means of Origin 9.1 software, using four different AFM images for each sample. The samples are prepared by drop-casting $MoSe_2$ flakes, graphene and SWCNT dispersions onto mica sheets (G250-1, Agar Scientific Ltd., Essex, U.K.) and drying them under vacuum.

Raman spectroscopy measurements are carried out using a Renishaw microRaman invia 1000 with a 50× objective, an excitation wavelength of 532 nm and an incident power of 1 mW. For each sample, 50 spectra are collected. The samples are prepared by drop casting $MoSe_2$ flakes, graphene flakes and SWCNT dispersions on $Si/SiO_2$ substrates and drying them under vacuum. The spectra are fitted with Lorentzian functions. Statistical analysis is carried out by means of OriginPro 9.1 software.

The crystal structure is characterized by XRD using a PANalytical Empyrean with Cu $K_\alpha$ radiation. The samples for XRD are prepared by drop-casting $MoSe_2$ flakes, graphene flakes and SWCNT dispersions onto $Si/SiO_2$ substrates and drying them under vacuum.

X-ray photoelectron spectroscopy characterization is carried out on a Kratos Axis UltraDLD spectrometer, using a monochromatic Al Kα source (15 kV, 20 mA). The spectra are taken over an area of 300 µm × 700 µm. Wide scans are collected with a constant pass energy of 160 eV and an energy step of 1 eV. High-resolution spectra are acquired at a constant pass energy of 10 eV and an energy step of 0.1 eV. The binding energy scale is referred to the C 1s peak at 284.8 eV. The spectra are analysed using CasaXPS software (version 2.3.17). The fitting of the spectra is performed by using a linear background and Voigt profiles. The samples are prepared by drop-casting $MoSe_2$ flakes, graphene flakes and SWCNT dispersions onto $Si/SiO_2$ substrates (LDB Technologies Ltd) and drying them under vacuum.

Optical absorption spectroscopy measurements are carried out using a Cary Varian 6000i UVvis-NIR spectrometer with a quartz glass cuvette with a path length of 1 cm. $MoSe_2$ flakes, graphene flakes and SWCNT dispersions are characterized as-produced. The corresponding solvent baselines are subtracted.

*Electrode fabrication*

MoSe$_2$ flakes are deposited on GC sheets (Sigma Aldrich) (GC/MoSe$_2$) by drop-casting the as-produced MoSe$_2$ flake dispersions (mass loading of 2 mg cm$^{-2}$). Graphene flakes and SWCNTs are deposited onto nylon membranes with a pore size of 0.2 μm (Whatman® membrane filters nylon, Sigma Aldrich) through a vacuum filtration process (mass loading of 2 mg cm$^{-2}$, electrode area of 3.14 cm$^2$). Hybrid electrodes are fabricated by depositing graphene flakes or SWCNTs and MoSe$_2$ flakes (graphene/MoSe$_2$ and SWCNT/MoSe$_2$) on nylon membranes with a pore size of 0.2 μm through sequential vacuum filtration processes (mass loading of 2 mg cm$^{-2}$ for both graphene and SWCNTs, and 2 or 5 mg cm$^{-2}$ for MoSe$_2$ flakes). The electrodes are dried overnight at room temperature before their electrochemical characterization. Thermal treatment of GC/MoSe$_2$ and SWCNT/MoSe$_2$ is carried out in a quartz tube (with a length of 120 cm and an inner diameter of 25 mm) which is passed through a three zone split furnace (PSC 12/--/600H, Lenton, UK). The electrodes are heated at 600, 700 or 800 °C (with a ramp of 12 °C/min) for 5 hours under a 100 sccm flow of Ar/H$_2$ (90/10%). Gas flows are controlled upstream by an array of mass flow controllers (1479A, mks, USA). Finally, the oven is cooled down to room temperature. Chemical treatment of SWCNTs/MoSe$_2$ is obtained by bathing them in 5 ml of n-butyllithium (Sigma Aldrich) in a sealed vial at room temperature in an N$_2$ atmosphere. After 12 hours, the electrodes are washed with deionized water to remove any remaining Li that is still present in the form of lithium cations (Li$^+$) then cleaned with IPA and dried with compressed N$_2$ gas.

*Electrode characterization*

Scanning electron microscopy analysis is performed with a field-emission scanning electron microscope FE-SEM (Jeol JSM-7500 FA). The acceleration voltage is set to 5kV. Images are collected using a secondary electron sensor for LEI images and the in-lens sensor for SEI images. Energy-dispersive X-ray spectroscopy images are acquired at 5kV by a silicon drift detector (Oxford Instruments X-max 80) with an 80mm$^2$ window. The EDX analysis is performed using Oxford Instrument AZtec 3.1 software.

Atomic force microscopy images are taken using the same setup as that of material characterization. Height profiles are processed by using JPK Data Processing software (JPK Instruments, Germany) and the data are analyzed with OriginPro 9.1 software. In addition to the graphene, SWCNTs, graphene/MoSe$_2$ and SWCNTs/MoSe$_2$ electrodes, samples of MoSe$_2$ flake films, produced by drop-casting MoSe$_2$ flake dispersions on Si/SiO$_2$ (mass loading of 2 mg cm$^{-2}$), are also imaged before and after thermal annealing in Ar/H$_2$ (90/10%).

Raman spectroscopy measurements are carried out using the same setup as that of material characterization. The analysis is performed on graphene/MoSe$_2$ and SWCNTs/MoSe$_2$ with a mass loading of 2 mg cm$^{-2}$ for each material (see Electrode fabrication section).

X-ray photoelectron spectroscopy characterization is carried out using the setup and analysis software described in the Material characterization section. The analysis is performed on graphene/MoSe$_2$ and SWCNTs/MoSe$_2$ with a mass loading of 2 mg cm$^{-2}$ for each material (see Electrode fabrication section). Measurements are also taken on MoSe$_2$ flake-based films, which are produced by drop-casting MoSe$_2$ flake dispersions onto Si/SiO$_2$ (mass loading of 2 mg cm$^{-2}$), before and after thermal annealing in Ar/H$_2$ (90/10%) or chemical bathing in 5 ml of n-butyllithium.

Electrochemical measurements on the as-prepared electrodes are carried out at room temperature in a flat-bottom fused silica cell under a three-electrode configuration using a CompactStat potentiostat/galvanostat station (Ivium), controlled via Ivium's own IviumSoft. A Pt wire is used as the counter-electrode and saturated KCl Ag/AgCl is used as the reference electrode. Measurements are carried out in 200 mL of 0.5 M H$_2$SO$_4$ (99.999% purity, Sigma Aldrich) (pH 1) or 1 M KOH ($\geq$ 85% purity, ACS reagent, pellets, Sigma Aldrich). Oxygen is purged from electrolyte by flowing N$_2$ gas throughout the liquid volume using a porous frit for 30 minutes before starting the measurements. A constant N$_2$ flow is then maintained for the whole duration of the experiments, to avoid re-dissolution of molecular oxygen in the electrolyte. The potential difference between the working electrode and the Ag/AgCl reference electrode is converted to the reversible hydrogen electrode (RHE) scale via the Nernst equation: $E_{RHE} = E_{Ag/AgCl} + 0.059 pH + E^0_{Ag/AgCl}$, in which $E_{RHE}$ is the converted potential versus RHE, $E_{Ag/AgCl}$ is the experimental potential measured against the Ag/AgCl reference electrode, and $E^0_{Ag/AgCl}$ is the standard potential of Ag/AgCl at 25 °C (0.1976 V). Polarization curves are acquired at a scan rate of 5 mV s$^{-1}$. Polarization curves from all catalysts are *iR*-corrected, in which *i* is the current and the *R* is the series resistance that arises from the substrate and electrolyte resistances. *R* is measured by EIS at an open circuit potential and frequency of 10$^4$ Hz. Electrochemical impedance spectroscopy measurements for GC/MoSe$_2$ films before and after thermal treatment in H$_2$ atmosphere are acquired in the 0.1 Hz ÷ 100 kHz frequency range at 0.2 V *vs.* RHE (near the equilibrium potential of the electrodes) with an AC amplitude of 0.02 V.

The linear portions of the Tafel plots fit the Tafel equation η = *b**|log(j)| + *A*,[178,179] in which η is the overpotential with respect to the reversible hydrogen electrode potential (RHE), j is the current density, *b* is the Tafel slope and *A* is the intercept of the linear regression. The j$_0$ is the current density calculated from the Tafel equation by setting η to zero. Stability tests are carried out by chronoamperometry measurements (*j-t* curves), *i.e.,* by measuring the current in potentiostatic mode at a fixed overpotential in 0.5 M H$_2$SO$_4$ over time (200 min). The applied overpotential is varied between different electrodes in order to give a cathodic current density value of 30 mA cm$^{-2}$ at *t* = 0 for all cases.

**Acknowledgements**

This project has received funding from the European Union's Horizon 2020 research and innovation program under grant agreement No. 696656—GrapheneCore1. We thank Sergio Marras (Materials Characterization Facility at Istituto Italiano di Tecnologia) for his support in XRD data acquisition and analysis and the Electron Microscopy facility at Istituto Italiano di Tecnologia for the support in SEM/TEM data acquisition.

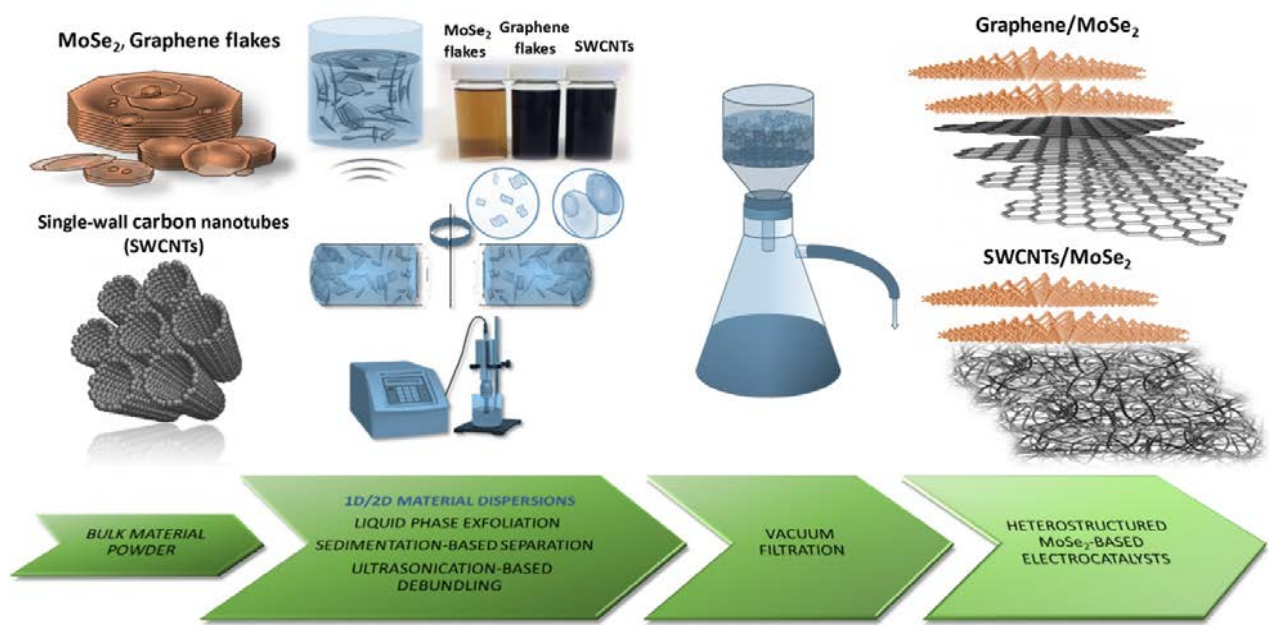

**Figure 1.** Solution-processed nanomaterials synthesis and MoSe$_2$-based electrocatalytic heterostructures fabrication.

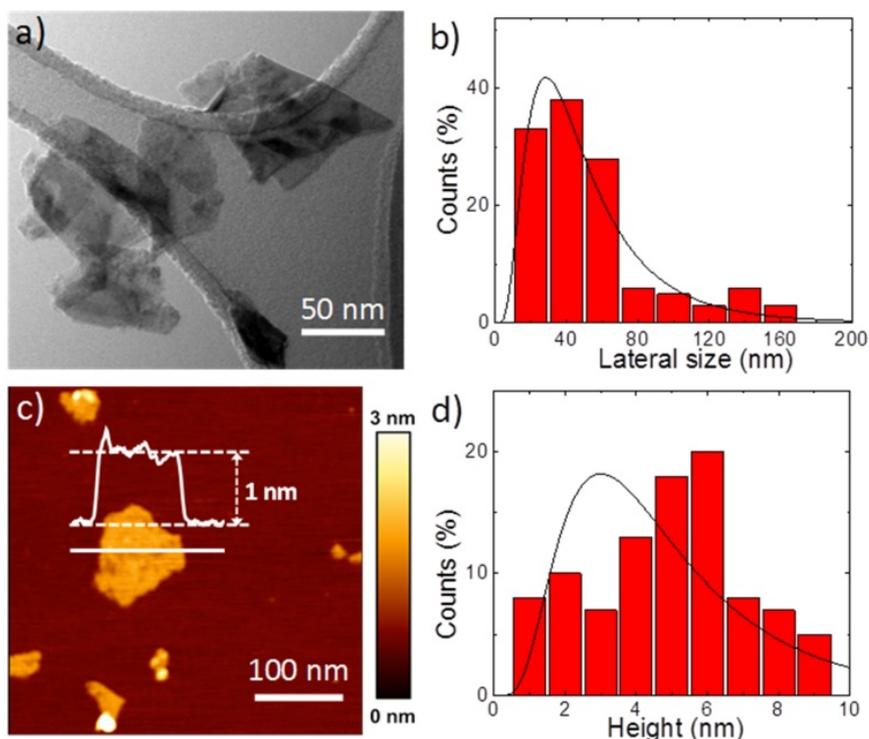

**Figure 2.** a) TEM images of the MoSe$_2$ flakes and b) the statistical analysis of their lateral dimension (calculated on 80 flakes). c) AFM images of MoSe$_2$ flakes deposited onto a mica sheet. The height profile of a representative flake is also shown (white line). d) Statistical analysis of the thickness of the MoSe$_2$ flakes (calculated on 80 flakes from different AFM images).

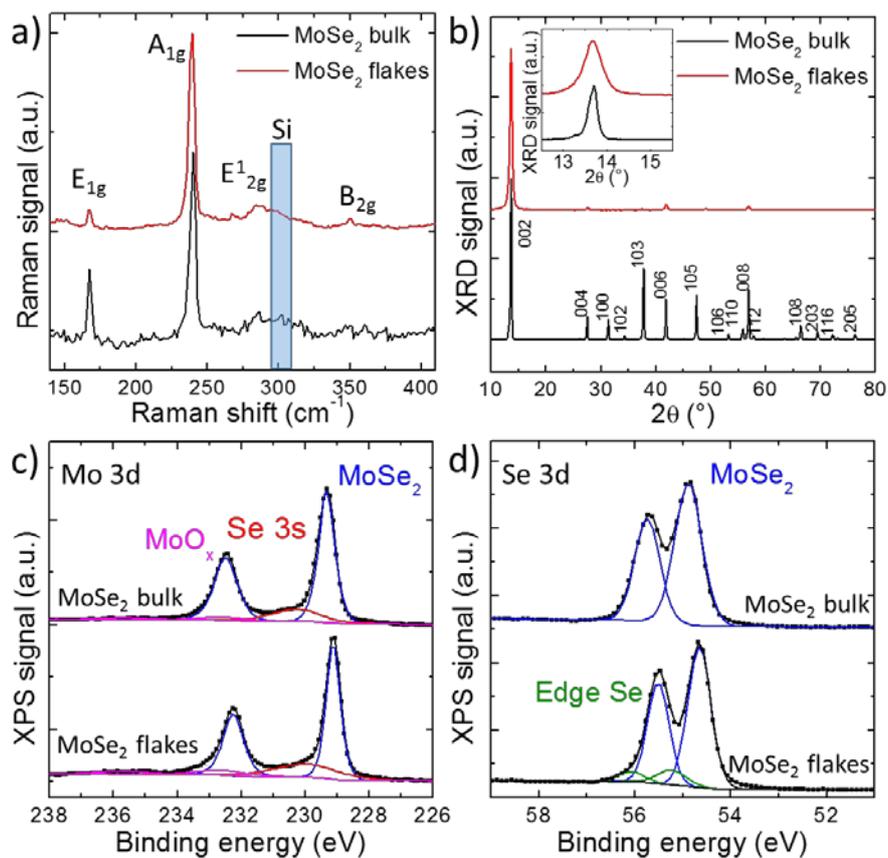

**Figure 3.** a) Raman spectra of the MoSe$_2$ bulk (black) and as-produced MoSe$_2$ flakes deposited on Si/SiO$_2$ substrates. The main peaks, *i.e.*, the in-plane modes E$_{1g}$, E$^1_{2g}$, and E$^2_{2g}$, the out-of-plane mode A$_{1g}$ and the breathing mode B$^1_{2g}$ are named in the graph. b) XRD spectra of the MoSe$_2$ bulk (black) and MoSe$_2$ flakes. The diffraction peaks of the hexagonal phase of MoSe$_2$ (2H-MoSe$_2$) are also indicated in the XRD spectra of the MoSe$_2$ bulk. The inset panel shows the broadening of the (002) peak of the MoSe$_2$ flakes. c) Mo 3d and d) Se 3d XPS spectra for MoSe$_2$ bulk (top curves) and MoSe$_2$ flakes (bottom curves). Their deconvolution is also shown, evidencing the bands ascribed to the: MoSe$_2$ (blue curves); Se 3s (red curve), which overlaps with the Mo 3d XPS spectrum; oxidized species (MoO$_x$) (magenta curves); edge (elemental) Se (green curves).

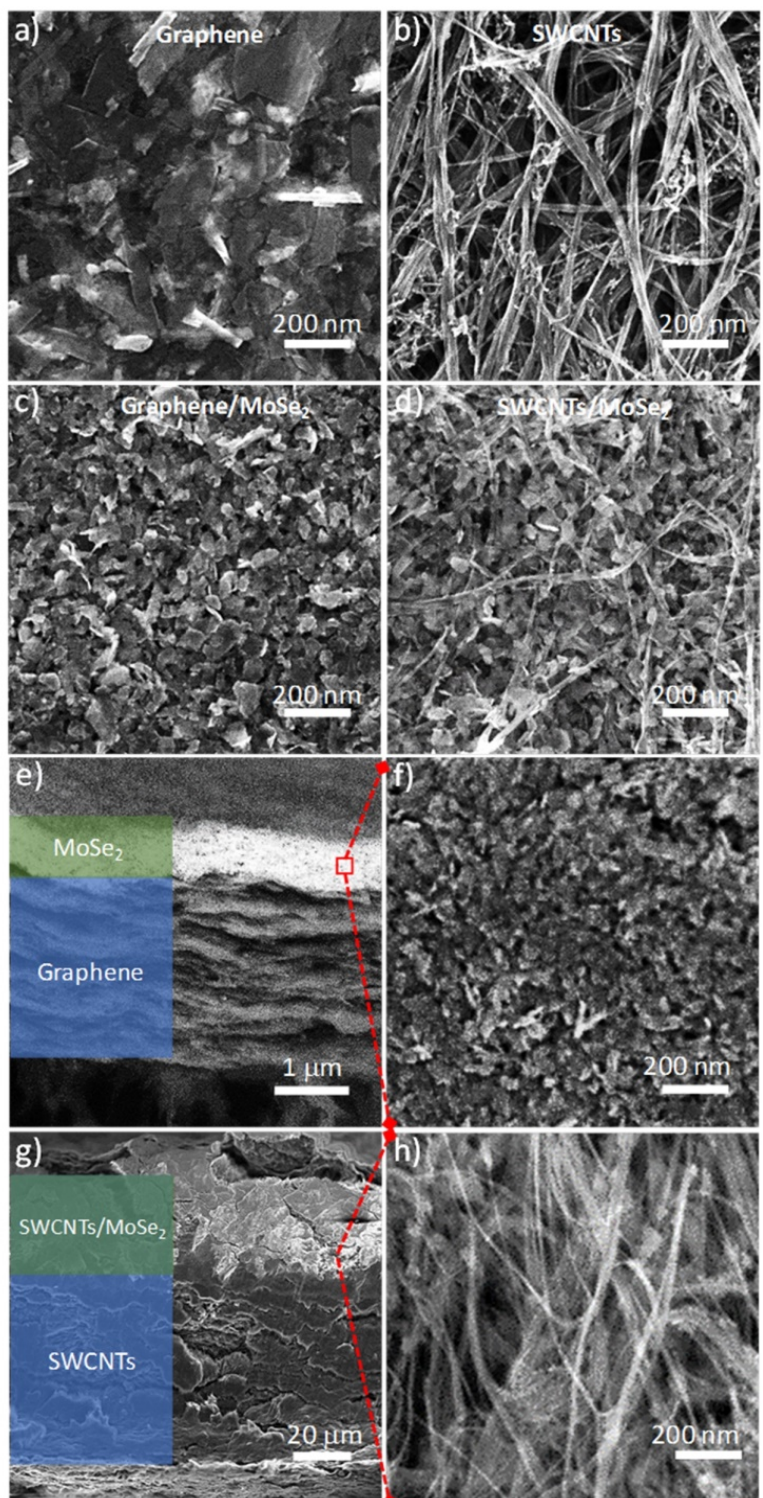

**Figure 4.** Top-view SEM images of a) graphene, b) SWCNTs, c) graphene/MoSe$_2$ and d) SWCNTs/MoSe$_2$. Cross-sectional SEM images of e-f) graphene/MoSe$_2$ and g-h) SWCNTs/MoSe$_2$. Panels f) and h) resolve the structures of the top-layers for the corresponding hybrid electrodes. The material mass loading is 2 mg cm$^{-2}$.

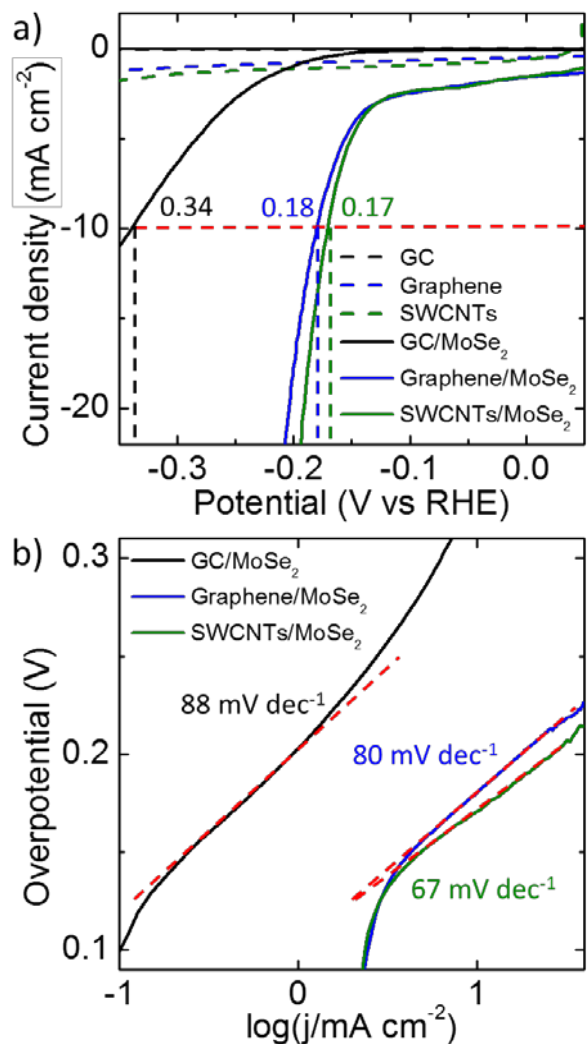

**Figure 5.** a) Polarization curves of GC/MoSe$_2$ (solid black line), graphene/MoSe$_2$ (solid blue line), SWCNTs/MoSe$_2$ (solid green line) in 0.5 M H$_2$SO$_4$. Polarization curves of GC (dashed black line), graphene (dashed blue line) and SWCNTs (dashed green line) are shown for comparison. b) Tafel plots of the GC/MoSe$_2$ (solid black line), graphene/MoSe$_2$ (solid blue line) and SWCNTs/MoSe$_2$ (solid green line). Linear fits (dashed red lines) and the corresponding Tafel slope values are reported.

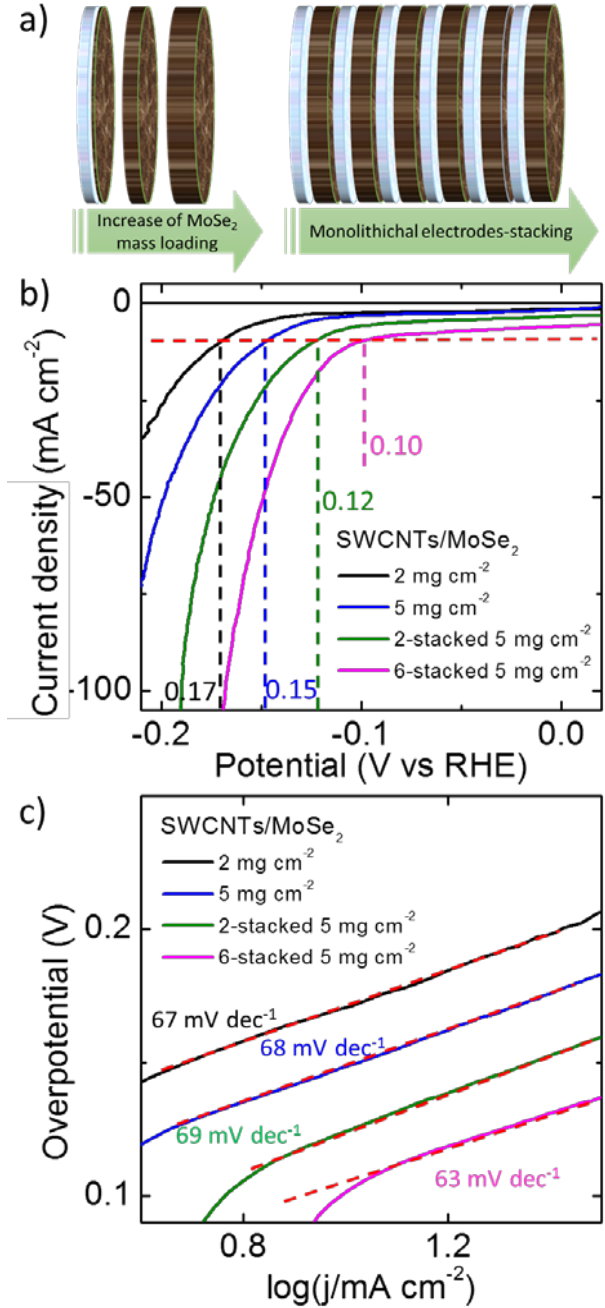

**Figure 6.** a) Sketch of the electrodes obtained by increasing the mass loading of the MoSe$_2$ flake and by the monolithical stacking of different electrodes. b) Polarization curves of SWCNT/MoSe$_2$ with a MoSe$_2$ flake mass loading of 2 mg cm$^{-2}$ (solid black line ), 5 mg cm$^{-2}$, (solid blue line), and the electrode obtained by the monolithical stacking of 2 and 6 SWCNTs/MoSe$_2$ with a MoSe$_2$ flake mass loading of 5 mg cm$^{-2}$ (solid green and magenta lines, respectively) in 0.5 M H$_2$SO$_4$. b) Tafel plots of SWCNT/MoSe$_2$ with a MoSe$_2$ flake mass loading of 2 mg cm$^{-2}$ (solid black line ), 5 mg cm$^{-2}$, (solid blue line), and the electrode obtained by the monolithical stacking of 2 and 6 SWCNTs/MoSe$_2$ with a MoSe$_2$ flake mass loading of 5 mg cm$^{-2}$ (solid green and magenta lines, respectively). Linear fits (dashed red lines) and the corresponding Tafel slope values are reported.

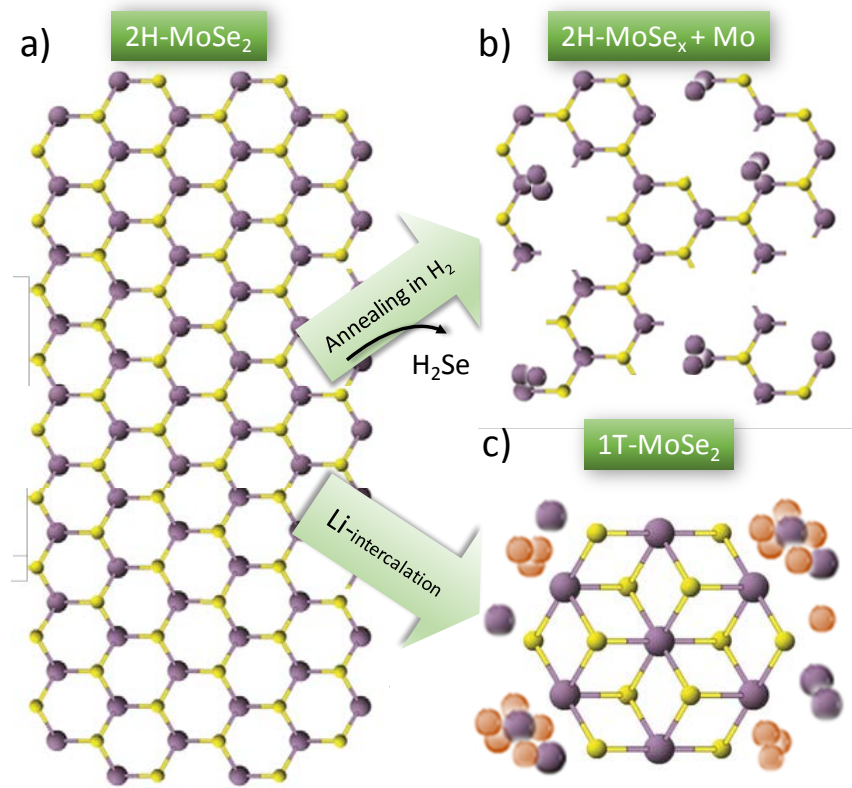

**Figure 7.** Schematic illustration of the treatment of MoSe$_2$ flakes for increasing their HER-electrocatalytic activity. a) As-produced 2H-MoSe$_2$ flake; b) Se-vacancy engineered 2H-MoSe$_2$ flake produced by thermo-induced flake texturization in an H$_2$ environment; c) 1T-MoSe$_2$ flake, MoO$_x$ and elemental atoms produced by *in situ* semiconducting-to-metallic phase conversion (as obtained by chemical bathing in n-butyllithium). Atom colour code: purple, Mo; yellow, Se.

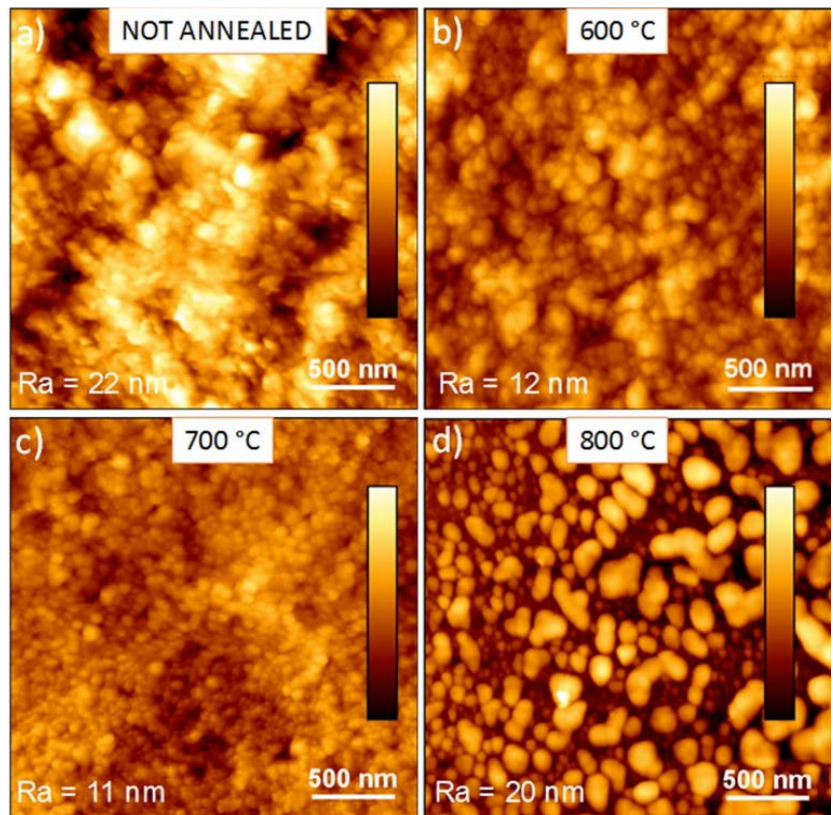

**Figure 8.** AFM images of MoSe$_2$ flake films deposited onto an Si substrate. a) untreated sample; a-c) samples annealed at 600 °C (panel b), 700 °C (panel c) and 800 °C (panel d) in Ar/H$_2$ (90/10%) for 5 h. The z-scale bar is 145 nm.

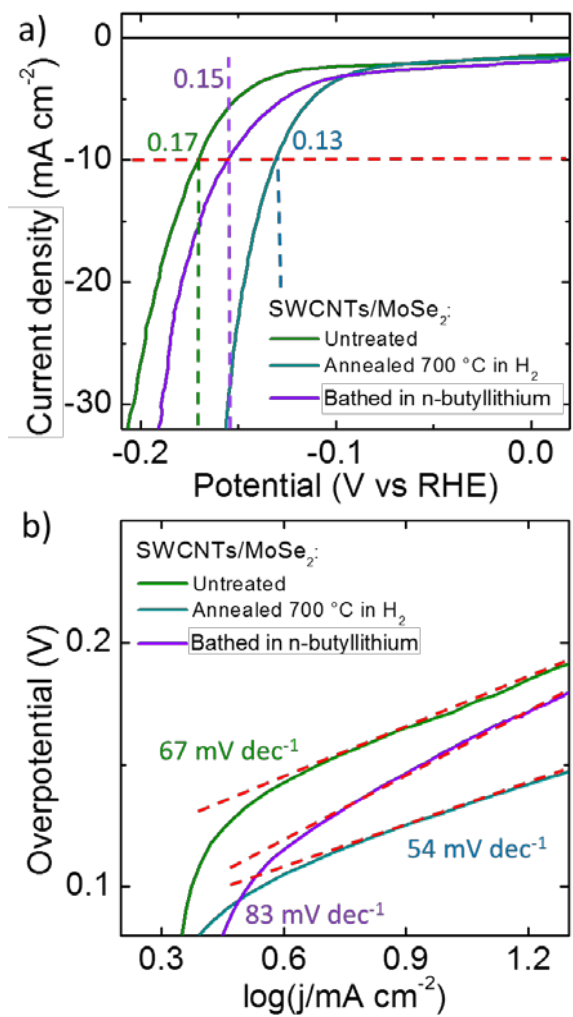

**Figure 9.** a) Polarization curves of untreated SWCNTs/MoSe$_2$ (solid green line), SWCNTs/MoSe$_2$ annealed at 700 °C in Ar/H$_2$ (90/10%) for 5 h (solid dark cyan line), SWCNTs/MoSe$_2$ chemically treated in n-butyllithium for 12 h (solid violet line) in 0.5 M H$_2$SO$_4$. b) Tafel plots of the SWCNTs/MoSe$_2$ (solid green line ), SWCNTs/MoSe$_2$ annealed at 700 °C in Ar/H$_2$ (90/10%) for 5 h (solid dark cyan line), SWCNTs/MoSe$_2$ bathed in n-butyllithium for 12 h (solid violet line). Linear fits (dashed red lines) and the corresponding Tafel slope values are reported.

**Thermo-induced texturization, chemically induced material phase conversion and the monolithic stacked assembly of solution-processed MoSe$_2$-based heterostructures provide advanced tools that are required to efficiently produce electrochemical hydrogen.**

**Keyworkds:** two-dimensional (2D) materials, molybdenum diselenide (MoSe$_2$), hydrogen evolution reaction (HER), graphene, carbon nanotubes (CNTs), liquid phase exfoliation (LPE)

**Engineered MoSe$_2$-based heterostructures for efficient electrochemical hydrogen evolution reaction**

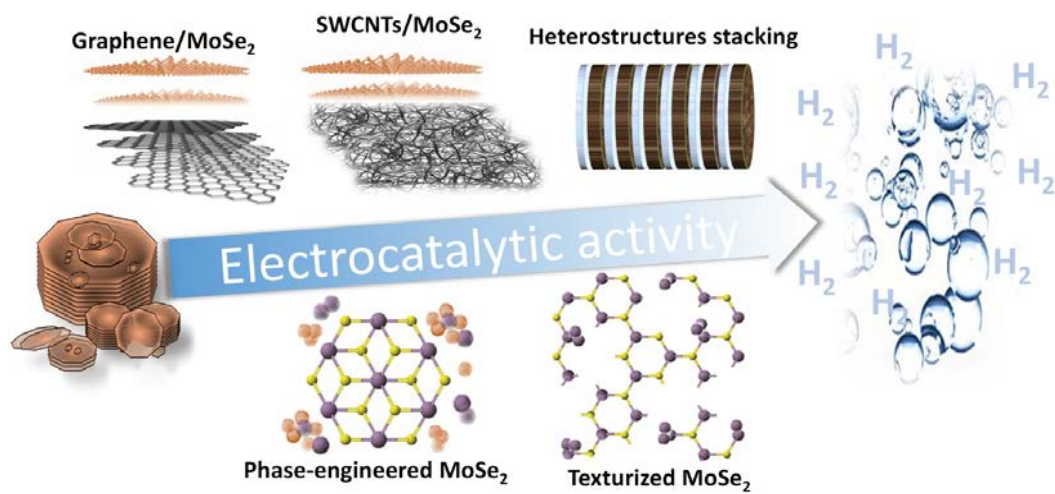

# Supporting Information

**Engineered MoSe₂-based heterostructures for efficient electrochemical hydrogen evolution reaction**

*Leyla Najafi, Sebastiano Bellani, Reinier Oropesa-Nuñez, Alberto Ansaldo, Mirko Prato, Antonio Esau Del Rio Castillo and Francesco Bonaccorso*[*]

**Morphological characterization of the MoSe₂ flakes**

**Figure S1**a shows the transmission electron microscopy (TEM) image of a single MoSe₂ flake, while Figure S1b reports the corresponding TEM-selected area electron diffraction (TEM-SAED), whose sharp ring-and-dot pattern indicates the polycrystalline nature of the flake, as is also evidenced by the XRD and Raman characterization in the main text (Figures 3a,b). Figure S1c reports the AFM image of MoSe₂ flakes with a thickness superior to those reported in the main text (~1 nm) (Figure 2c). Figure S1d reports the AFM height profile of a representative horizontal section of Figure S1c (white dashed line), in which two superimposed flakes with a thickness less than 3 nm are imaged.

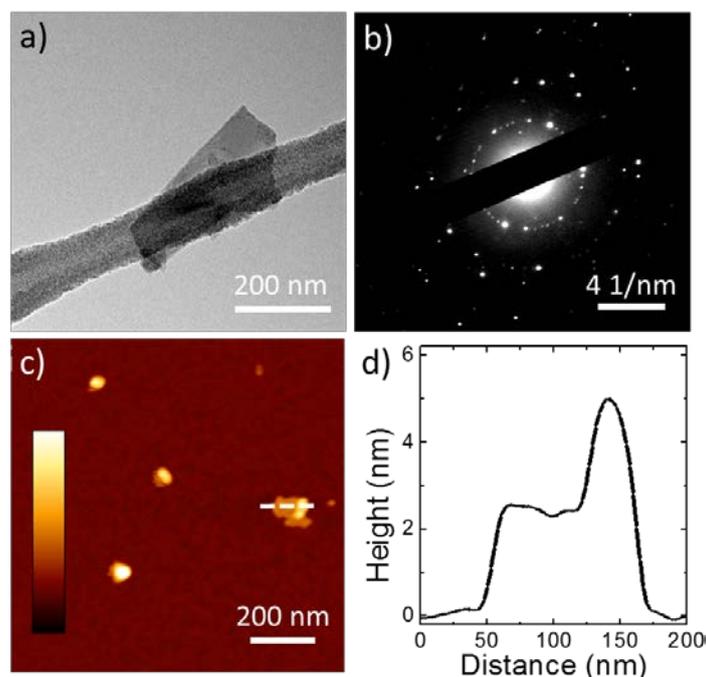

**Figure S1.** a) TEM image of a single MoSe₂ flake and b) its corresponding TEM-selected area electron diffraction (TEM-SAED). c) AFM image of few-layered MoSe₂ flakes (z-scale bar is 8 nm). d) The AFM Height profile of superimposed MoSe₂ flakes taken along the horizontal section is indicated by a white dashed line in panel c).

**Optical absorption spectroscopy of MoSe$_2$ flakes**

**Figure S2** reports the optical absorption spectroscopy (OAS) of MoSe$_2$ flake dispersion in IPA. As discussed in the main text of the manuscript, the absorption peaks around 810 nm (1.53 eV) and 708 nm (1.75 eV) correspond to the A and B excitonic peaks. These peaks arise from the direct inter-band transitions at the K-point of the Brillouin zone of the 2H-phase MoSe$_2$,[1-4] as originated from the energy split of the valence-band that was formed from the Mo atom[5,6] which, in turn, was caused by the interlayer coupling[5-7] and spin-orbit interaction effects[6,7] in few-layered MoSe$_2$ flakes. The shoulder in the absorption spectra around ~410 nm is attributed to the C and D inter-band transitions between the density of state peaks in the valence and conduction bands of the 2H-phase of MoSe$_2$.[8,9]

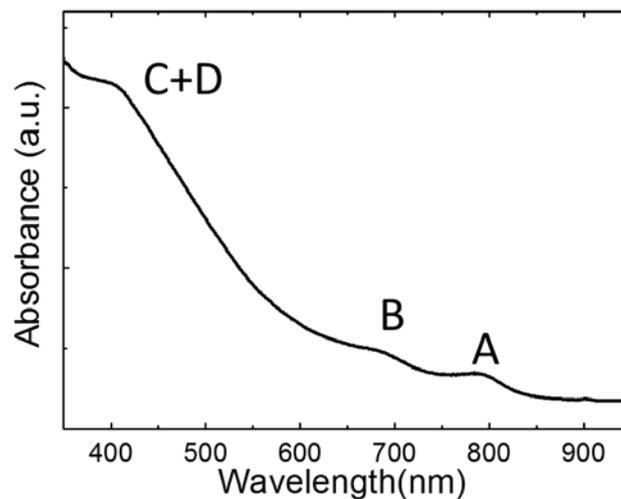

**Figure S2.** UV-Vis absorption spectra of MoSe$_2$ flake dispersions in IPA. The excitonic peaks (A and B) and the inter-band transitions (C and D) are also indicated.

**Statistical Raman analysis of MoSe$_2$ flakes**

**Figure S3** shows the statistical Raman analysis of the Pos (A$_{1g}$) of MoSe$_2$ bulk (panel a) and the Pos(A$_{1g}$), Pos(E$^1_{2g}$) and I(A$_{1g}$)/I(E$_{2g}$) of MoSe$_2$ flakes (panel b, c and d, respectively). As discussed in the main text of the manuscript, the mode A$_{1g}$ is located at ~241 cm$^{-1}$ for the MoSe$_2$ bulk, while it is red-shifted to ~239 cm$^{-1}$ for the MoSe$_2$ flakes, which is in agreement with the softening of the vibrational mode that is accompanied by the reduction in layer thickness.[10-14] The intensity ratio between the A$_{1g}$ and E$^1_{2g}$ modes (I(A$_{1g}$)/I(E$^1_{2g}$)) for MoSe$_2$ flakes is ∼21. This values is consistent with those already reported for few-layered MoSe$_2$ flakes.[15,16]

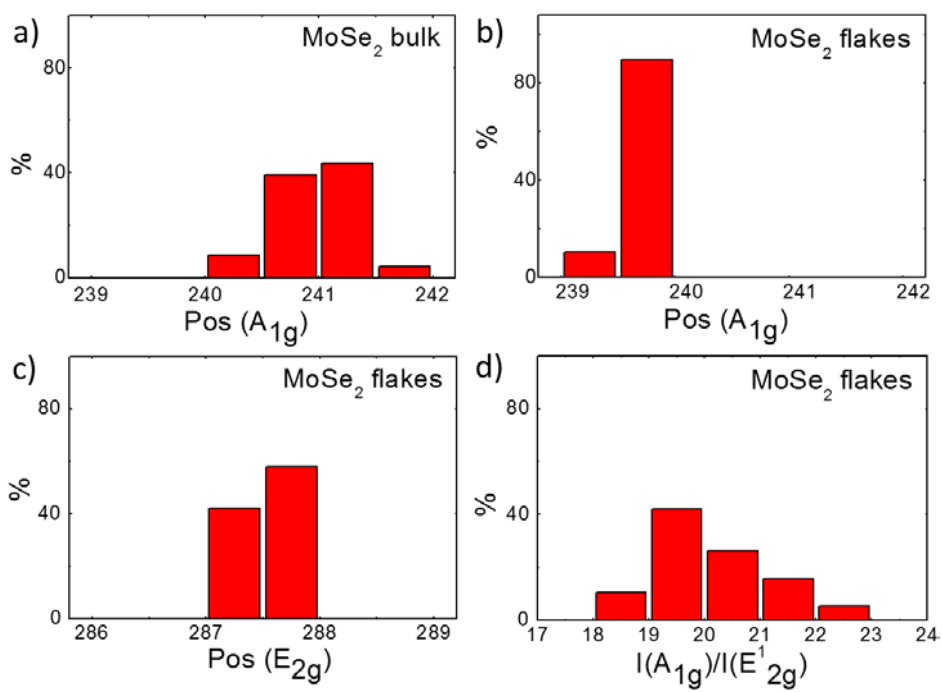

**Figure S3.** Statistical Raman analysis of: a) Pos (A$_{1g}$) of MoSe$_2$ bulk; b) Pos(A$_{1g}$), c) Pos(E$_{2g}$) and I(A$_{1g}$)/I(E$_{2g}$) of MoSe$_2$ flakes.

**Morphological, optical, structural and chemical characterization of graphene flakes**

The morphology of the as-produced graphene flakes is characterized by means of TEM and atomic force microscopy (AFM). **Figure S4**a shows a representative TEM image of graphene flakes, which have an irregular shape and rippled morphology. Statistical TEM analysis of the flakes lateral size (Figure S4b) indicates values that are distributed in the range of 200-1500 nm and an average value of ~450 nm. Figure S4c shows a representative AFM image of graphene flakes. The main thickness distribution is in the range of 0.5-4.0 nm (Figure S4d), but a few thicker flakes are also present (*i.e.*, >5 nm).

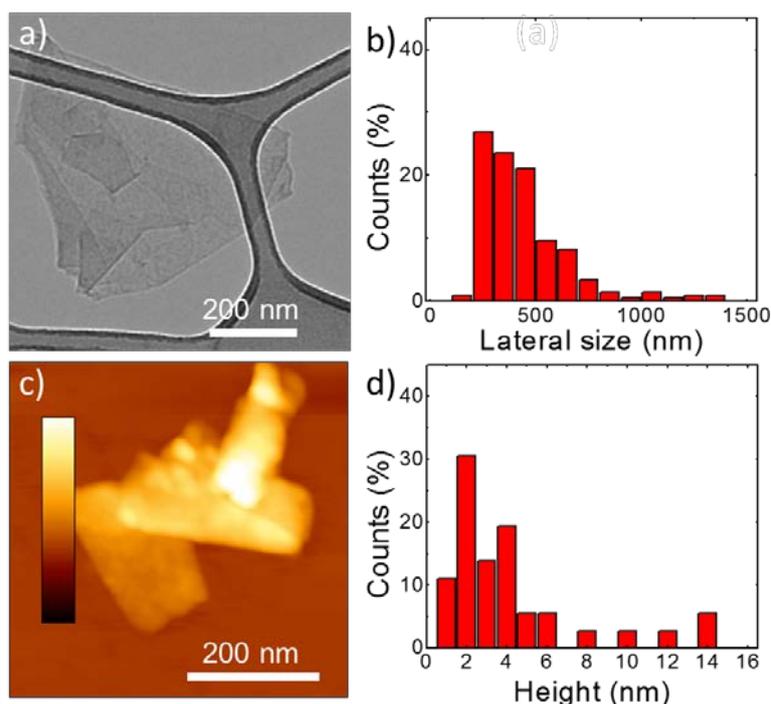

**Figure S4.** a) TEM images of the as-produced graphene flakes and b) TEM statistical analysis of their lateral dimension. c) AFM images of the as-produced graphene flakes (z-scale bar is 8 nm) and d) AFM statistical analysis of their lateral dimension (calculated on different AFM images).

**Figure S5** reports the optical absorption spectroscopy (OAS) measurement of the as-produced graphene flake dispersion in N-Methyl2Pyrrolidone (NMP), showing a peak at ~265 nm. This peak is a signature of the van Hove singularity in the graphene density of states.[17]

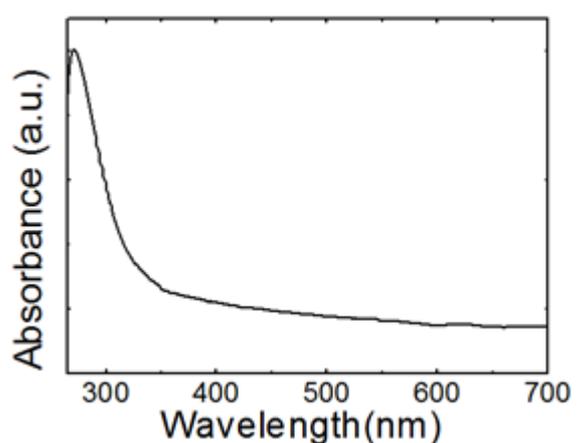

**Figure S5.** Absorbance spectrum of the as-produced graphene flake dispersion in NMP.

Raman spectroscopy is used to evaluate the structural properties of the graphene flakes. A typical Raman spectrum of defect-free graphene shows, as signature, G and D peaks.[18] The G peak corresponds to the $E_{2g}$ phonon at the Brillouin zone center.[19] The D peak is due to the breathing modes of the $sp^2$ rings and requires a defect for its activation by double resonance.[18-21] The 2D peak is the second order of the D peak;[22] which appears as a single peak in monolayer graphene, but splits in four in bi-layer graphene, reflecting the evolution of the band structure.[18] The 2D peak is always seen, even in the absence of a D peak, since no defects are required for the activation of two phonons with the same momentum and one is backscattered from the other.[22] Double resonance can also happen as an intra-valley process, *i.e.,* connecting two points belonging to the same cone around K or K'.[22] This process gives rise to the D' peak for defective graphene.[22] D+D' is the combination mode of D and D', while 2D' is the second order of the D'.[22] As in the case of 2D, 2D' is always seen even when the D' peak is not present.[22] **Figure S6**a reports a representative Raman spectrum of the as-produced graphene flakes, showing all the aforementioned bands. The statistical analysis of the G peak position (Pos(G)) (Figure S6b), the full width half maximum of G (FWHM(G)) (Figure S6c), the 2D peak position (Pos(2D)) (Figure S6d), the full width half maximum of 2D (FWHM(2D)) (Figure S6e), the intensity ratio between the 2D and G peaks (I(2D)/I(G)) (Figure S6f) and the intensity ratio between the D and G peaks (I(D)/(IG)) (Figure S6g) give quantitative information on the morphology of the graphene flakes. In particular, the Pos(2D) peaks at ~2700 cm$^{-1}$ (Figure S6d) while the FWHM(2D) ranges from 60 to 75 cm$^{-1}$ (Figure S6e). These values are ascribed to the presence of few-layered graphene (FLG).[18,23,24] The I(2D)/I(G) varies from 0.6 to 1.2 (Figure S6f), as is to be expected from a combination of single-layered graphene (SLG) and FLG.[18,25] The presence of D and D' indicates the defective nature of the graphene flakes.[22,26-28] Previous studies on graphene flakes produced by LPE have shown that these defects are predominantly located at the edges, while the basal plane of the flakes is defect-free.[27,28] This is demonstrated by the fact that there is no correlation between I(D)/I(G) and FWHM(G).[27,28] Figure S6g shows the statistical analysis of I(D)/I(G), which varies between 0.3 and 0.7, while Figure S6h does not show any correlation between I(D)/I(G) and FWHM(G), which is in agreement with data in literature,[27,28] thus the basal planes of the as produced graphene flakes are defect-free.[26-28] Overall, the microscopic (TEM, AFM) and spectroscopic (Raman) characterization indicates that the sample is mostly composed by a combination of sub-micrometric single-layered graphene (SLG) and few-layered graphene (FLG) flakes.

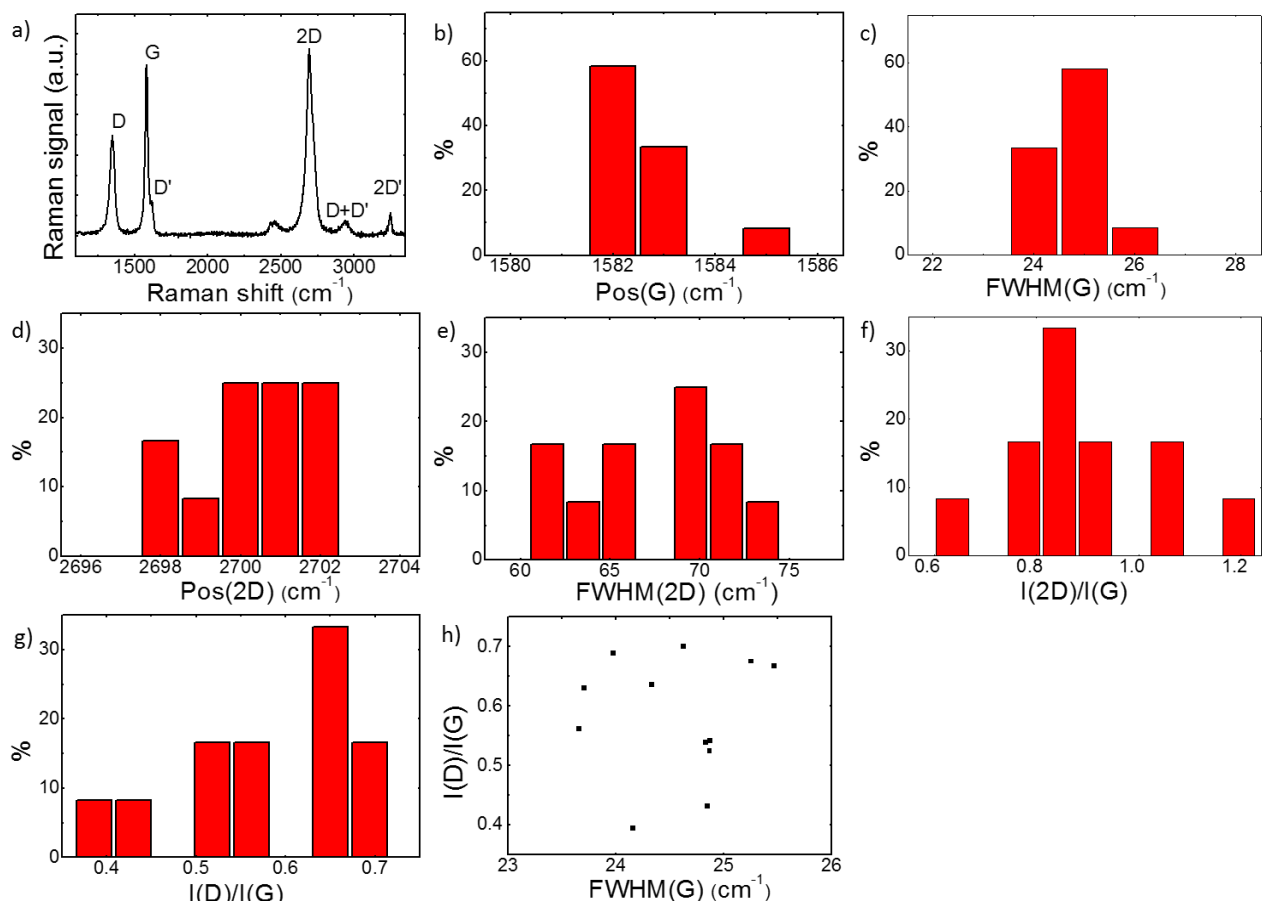

**Figure S6.** a) Representative Raman spectrum of the as-produced SLG/FLG by LPE in NMP. The D, G, D', 2D, D+D' and 2D' bands are also noted. b) Statistical Raman analysis of the Pos (G), c) FWHM(G), d) Pos(2D), e) FWHM(2D), f) I(2D)/I(G), g) I(D)/I(G) and h) I(D)/I(G) *vs.* FWHM(G) plot.

**Morphological, optical, structural and chemical characterization of SWCNTs**

**Figures S7**a,b shows representative TEM images of bundled SWCNTs. The length of the SWCNTs is between 5-30 µm (Figure S7a), which is in agreement with their datasheet that was provided by the supplier company (Cheap Tubes, see Experimental section in the main text for details). Amorphous carbon is not evident in the highest magnification of Figure S7b,

which is consistent with the declared quality of the SWCNTs (amorphous carbon content < 3%).

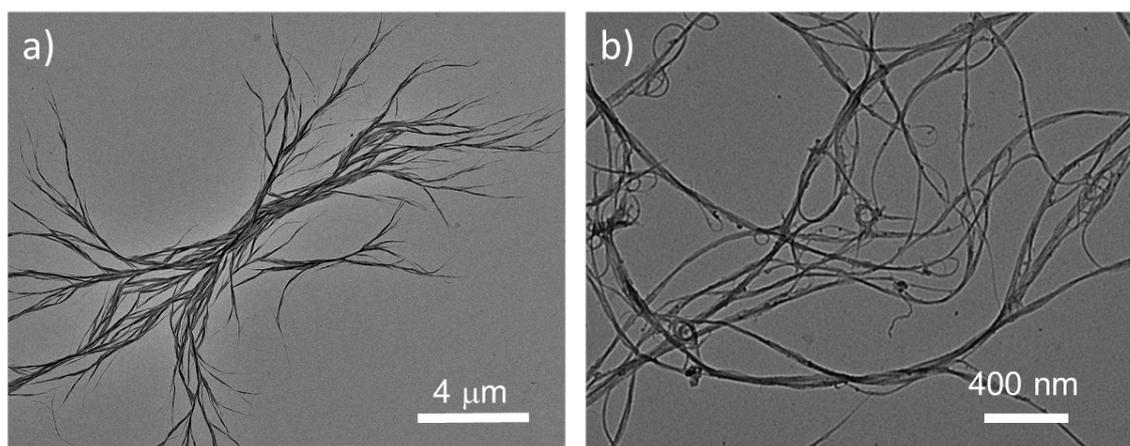

**Figure S7.** TEM images of bundled SWCNTs at different magnifications: scale bar of a) 4 µm and b) 400 nm.

**Figure S8**a reports the UV-Vis/NIR absorption spectrum of the SWCNT dispersion in NMP. The absorption spectrum of SWCNTs is characterized by a series of relatively sharp inter-band transitions, at energies noted as $E_{11}$, $E_{22}$, etc., which are associated with van Hove singularities.[29] The latter are correlated with the electronic structure of SWCNTs. In fact, the quasi 1D nature of SWCNTs, and their semiconducting versus metallic character, cause the electronic density of states to have a series of sharp van Hove maxima at energies that depend on the tube diameter (d) and the chiral wrapping angle describing its construction from a graphene sheet.[29,30] The spectrum shown in Figure S8a is consistent with this expectation, and the first van Hove transitions ($E_{11}$) of the direct band gap semiconducting tubes fall in the wavelength range of 900-1300 nm, while their subsequent van Hove transitions ($E_{22}$) are located between 550 and 900 nm. The lowest energy van Hove transitions of the metallic SWCNTs also appear between 400 and 600 nm. All these transitions indicate that the diameter distribution is in the range of 0.8-1.2 nm.[30,31]

Raman spectroscopy is also a useful tool for studying the structure of SWCNTs. The main features of the Raman spectrum of SWCNTs are the radial breathing modes (RBMs),[32] and the D, $G^+$ and $G^-$ and 2D peaks. The Pos(RBM) is inversely related to the diameter of the SWCNT (d),[33-35] as calculated by Pos(RBM)= $C_1$/d +$C_2$. In this study, $C_1$=214.4 cm$^{-1}$ nm and $C_2$=18.7 cm$^{-1}$, which is in agreement with previous studies.[36] Raman spectroscopy also probes possible damage, *i.e.,* the presence of defects, via the *D* peak.[37] The *D* peak arises due to the breathing modes of the sp$^2$ rings, which require a defect for their activation by double resonance.[20,38] The $G^+$ and $G^-$ bands are located between 1500-1600 cm$^{-1}$. These originate from the longitudinal (LO) and tangential (TO) modes, respectively, and derive from splitting the $E_{2g}$ phonon of graphene at the Brillouin zone centre.[19,39,40] The positions of the $G^+$ and $G^-$ peaks, Pos($G^+$), and Pos($G^-$), are depend on the diameter and their separation increases when the diameter is decreased.[41,42] In metallic SWCNTs, the FWHM ($G^-$) is larger and the Pos($G^-$) is down-shifted with respect to the semiconducting SWCNTs.[32,43] Thus, a wide, low frequency $G^-$ peak is a characteristic of metallic SWCNTs. The Raman spectrum of our SWCNTs which was acquired at an excitation wavelength of 532 nm is shown in Figure S8b. The spectrum shape shows the presence of metallic SWCNTs, while the analysis of the Pos(RBM) indicates that d <1 nm. A weak D peak is also observed (*i.e*., I(*D*)/I(*G*) <0.15). This is a signature of the presence of a small number of defects.[37]

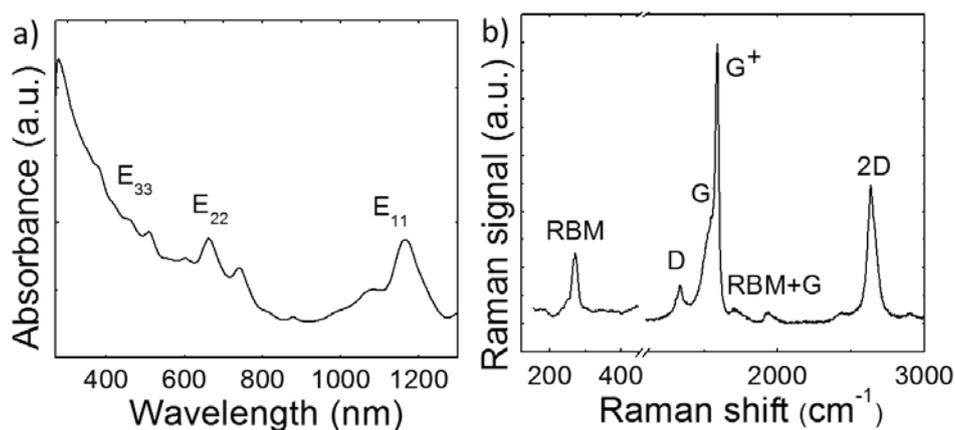

**Figure S8** a) UV-Vis/NIR absorption spectrum of SWCNT dispersion in NMP. The van Hove transitions are indicated. b) Raman spectrum of SWCNTs. The main Raman modes are labeled.

# AFM analysis of the graphene, SWCNTs, graphene/MoSe$_2$ and SWCNTs/MoSe$_2$ electrodes

**Figure S9** reports the AFM images of the electrodes surface, which display morphologies similar to those observed by SEM (see main text, Figure 4). The roughness average (Ra) values are ~46.2 nm and ~103 nm for graphene and SWCNT electrodes, respectively. These values decrease to ~21 nm and 70 nm, respectively, for the corresponding hybrid electrodes, indicating that the MoSe$_2$ flake deposition flattens the surface of the electrodes.

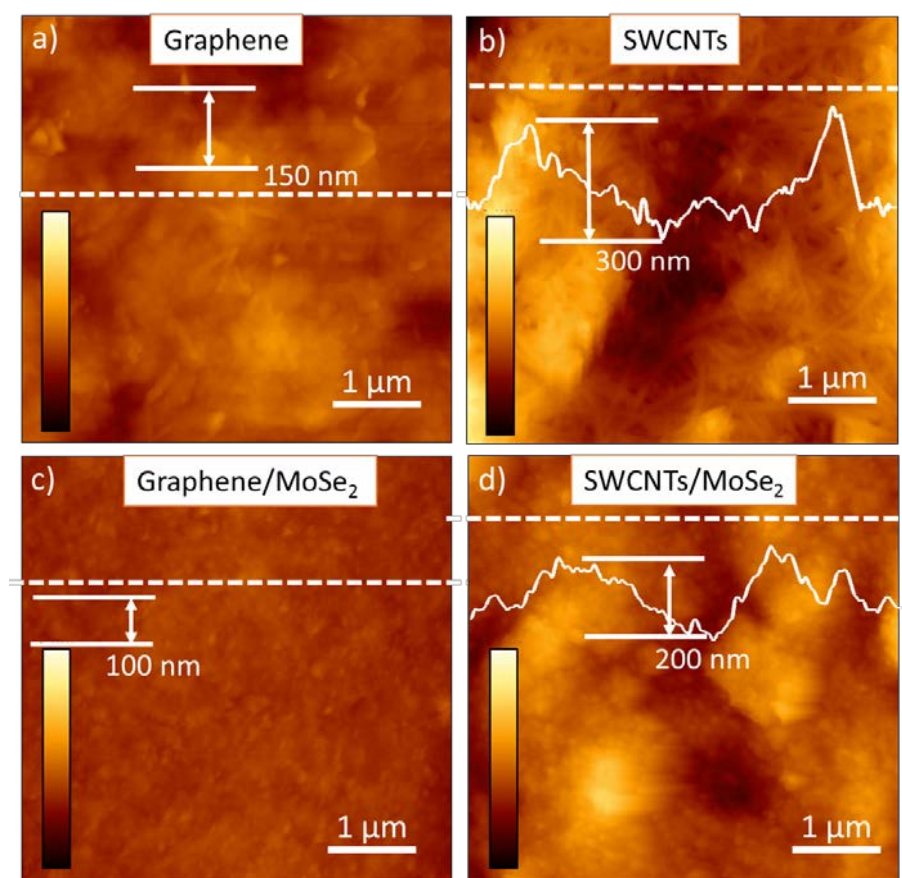

**Figure S9.** AFM images of a) graphene, b) SWCNTs, c) graphene/MoSe$_2$ and d) SWCNTs/MoSe$_2$. Height profiles along representative cross sections (white dashed lines) are also shown. The z-scale bar is 1 μm.

# Elemental energy-dispersive X-ray spectroscopy (EDX) analysis of the cross-sectional SEM images of SWCNTs/MoSe$_2$

A high-magnification cross-sectional SEM image of the SWCNTs/MoSe$_2$ (see main text, Figure 4h) reveals that the MoSe$_2$ flakes penetrate the SWCNT network. **Figure S10**a shows the EDX analysis (atom color code: yellow C; cyan Mo; violet Se) of a representative cross-sectional SEM image of SWCNTs/MoSe$_2$, while Figure S10b reports the corresponding mass spectrum. Figure S10c demonstrates a magnified region of Figure S10a, while Figures S10d-f display the corresponding EDX analysis for C, Mo and Se atoms, respectively. Notably, these results clearly highlight the presence of an interlayer of MoSe$_2$ flakes which penetrate the SWCNT network.

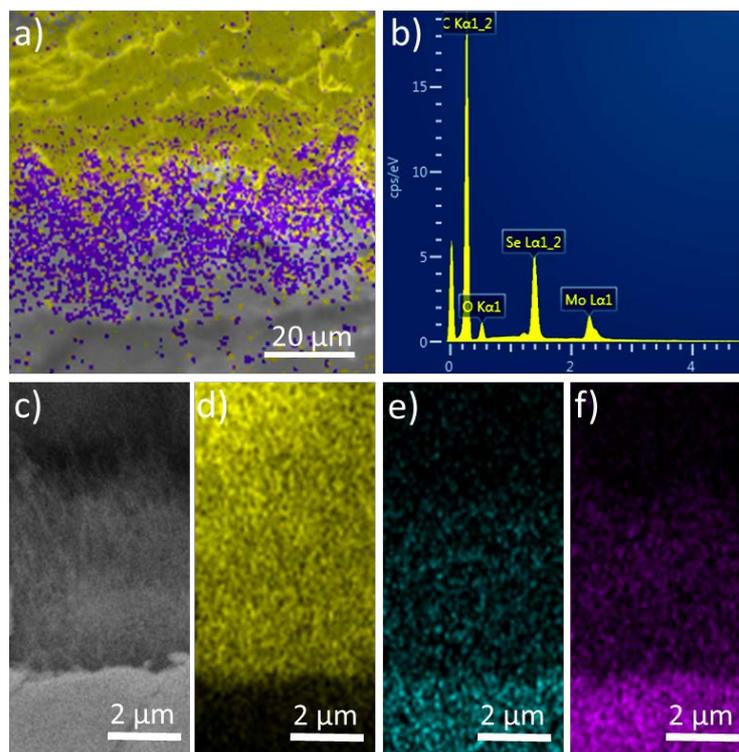

**Figure S10.** a) EDX analysis of a representative cross-sectional image of SWCNTs/MoSe$_2$ with b) the corresponding mass spectrum. c) Cross-sectional SEM images of a magnified region of the image of panel a), and the corresponding EDX analysis for d) C, e) Mo and f) Se atoms. Atom color code: yellow is C; cyan is Mo; violet is Se.

**Raman spectroscopy analysis of the graphene/MoSe$_2$ and SWCNTs/MoSe$_2$**

Raman spectroscopy measurements are carried out on the as-produced heterostructures (graphene/MoSe$_2$ and SWCNTs/MoSe$_2$, MoSe$_2$ flake mass loading of 2 mg cm$^{-2}$) in order to investigate the structural properties of the MoSe$_2$ flakes after their deposition on graphene and SWCNTs via vacuum filtration (see main text, Experimental Section). The Raman spectra of the graphene/MoSe$_2$ and SWCNTs/MoSe$_2$ in the spectral region of 140-410 cm$^{-1}$, where the Raman peaks of MoSe$_2$ flakes are located (see Figure 3a), are reported in **Figure S11**. The comparison with the Raman spectrum of the as-produced MoSe$_2$ flakes (as shown in the main text, Figure 3a) does not reveal any significant differences, which indicates that no structural modifications of the MoSe$_2$ flakes occur during their film deposition through the vacuum filtration of their dispersions.

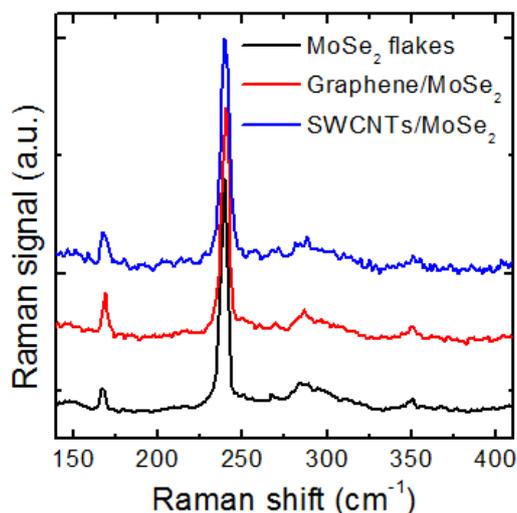

**Figure S11.** Raman spectra of MoSe$_2$ flakes (black) deposited on Si/SiO$_2$ substrates and the as-produced heterostructures (graphene/MoSe$_2$ flakes (red) and SWCNTs/MoSe$_2$ (blues)) (MoSe$_2$ flake mass loading of 2 mg cm$^{-2}$). The main peaks of the MoSe$_2$ flakes, *i.e.,* the in-plane modes E$_{1g}$, E$^1_{2g}$, and E$^2_{2g}$, the out-of-plane mode A$_{1g}$ and the breathing mode B$^1_{2g}$ are named in the graph.

**X-ray photoelectron spectroscopy (XPS) analysis of heterostructures**

**Figure S12** reports the XPS measurements of the graphene/MoSe$_2$ and SWCNTs/MoSe$_2$ surfaces, consisting of the MoSe$_2$ flake overlays (MoSe$_2$ flake mass loading of 2 mg cm$^{-2}$). Mo 3d XPS spectra (Figure S12a) show the two Mo 3d$_{5/2}$ and Mo 3d$_{3/2}$ of Mo(IV) states in MoSe$_2$ peaks[44-48] (see main text, Figure 3c) (located at: (229.2±0.2) eV and (232.3±0.2) eV for graphene/MoSe$_2$ ; (229.1±0.2) eV and (232.2±0.2) eV for SWCNTs/MoSe$_2$). The additional peaks (located at: (233.1±0.2) eV and (236.2±0.2) eV for graphene/MoSe$_2$; (232.9±0.2) eV and (236.0±0.2) eV for SWCNTs/MoSe$_2$) are assigned to the Mo(VI) state and are related to MoO$_3$ residues[49-51] (see main text, Figure 3c). The compositional analysis indicates that the percentage content (%c) of MoO$_3$ (defined as MoO$_3$/(MoO$_3$+MoSe$_2$)) is ~12% for graphene/MoSe$_2$ and ~17% for SWCNTs/MoSe$_2$. The percentage content (%c) of MoO$_3$ increase by 9% for graphene/MoSe$_2$ and 54% for SWCNTs/MoSe$_2$ with respect to that of the as-produced MoSe$_2$ flakes (~11%, see main text). Se 3d spectra (Figure S12b) show the Se 3d$_{5/2}$ and Se 3d$_{3/2}$ peaks of the diselenide moiety of MoSe$_2$[46-48,52-54] (see main text, Figure 3d) (located at: (54.8±0.2) eV and (55.6±0.2) eV for graphene/MoSe$_2$; (54.6±0.2) eV and (55.5±0.2) eV for SWCNTs/MoSe$_2$. For both the Mo 3d and the Se 3d spectra, a slight decrease in the binding energy (~0.1 eV for graphene/MoSe$_2$ and ~0.2 eV for SWCNTs/MoSe$_2$) is observed with respect to those of the as-produced MoSe$_2$ flakes (see main text, Fig 2c,d). These changes might be attributed to the μm-spatial range electrochemical coupling of the MoSe$_2$ flakes with the low-dimension carbon-based substrates.

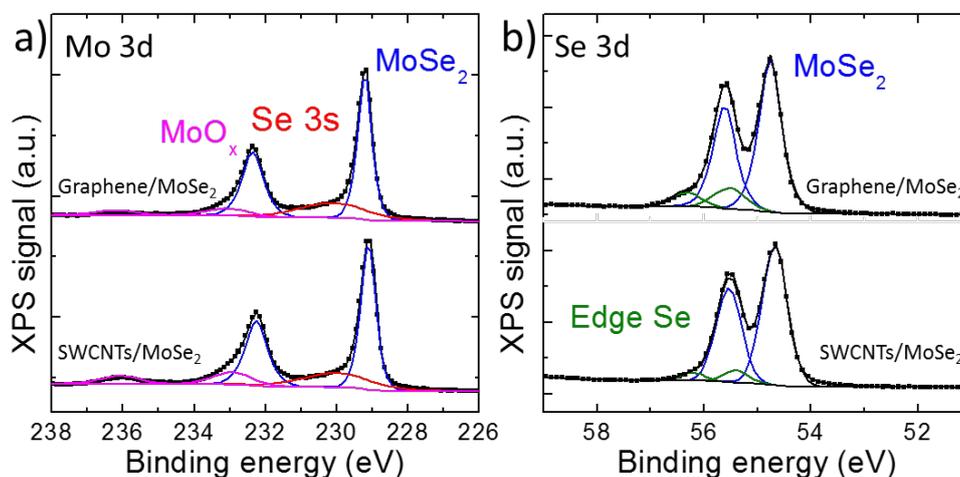

**Figure S12.** a) Mo 3d and b) Se 3d XPS spectra for graphene/MoSe$_2$ (top curves) and SWCNTs/MoSe$_2$ (bottom curves). Their deconvolution is also shown, evidencing the band attributed to: MoSe$_2$ (blue curves); Se 3s band (red curve), which overlaps with the Mo 3d XPS spectrum; oxidized species (MoO$_x$) (magenta curves); edge (elemental) Se (green curves).

**HER-electrocatalytic activity of the MoSe$_2$ flakes in acid *vs.* alkaline solutions**

**Figure S13** shows the comparison between the HER-electrocatalytic activity of the MoSe$_2$ flakes in acid (0.5 M H$_2$SO$_4$) and in alkaline (1 M KOH) solutions. As stated in the main text, the HER in acid solution is assumed to be proceeded by an initial discharge of the hydronium ion (H$_3$O$^+$) and the formation of hydrogen is intermediated, *i.e.*, atomic hydrogen adsorbed (H$_{ads}$) in the so-called Volmer step (H$_3$O$^+$ + $e^-$ → H$_{ads}$ + H$_2$O) is followed by either an electrochemical Heyrovsky step (H$_{ads}$ + H$_3$O$^+$ + $e^-$ → H$_2$ + H$_2$O) or a chemical Tafel recombination step (2H$_{ads}$ → H$_2$). However, in alkaline conditions, the H$_{ads}$ is formed by the discharge of $H_2O$ ($H_2O + e^- \rightarrow H_{ad} + OH^-$). Then, either the Heyrovsky step ($H_2O + H_{ads} + e^- \rightarrow H_2 + OH^-$) or the chemical Tafel recombination step ($2H_{ads} \rightarrow H_2$) occur. The overpotential *vs.* reversible hydrogen electrode (RHE) scale at a cathodic current density ($\eta_{10}$) of 10 mA cm$^{-2}$ is 0.34 V in acid solution (see also main text, Figure 5) and 0.37 V in alkaline solution. The $\eta_{10}$ observed in alkaline solution is higher than the acid one, which has been attributed to the high kinetic energy barrier of the initial $H_2O$ discharge and the $H_{ads}$ formation, as well as to the strong adsorption of the $OH^-$ that formed on the surfaces of the $MoSe_2$ flakes (and, in general, of the 2D-TMDs).[55,56]

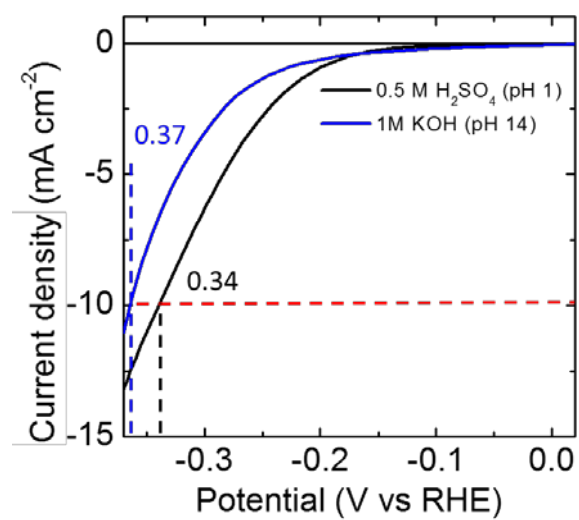

**Figure S13.** Polarization curves of GC/MoSe$_2$ in acid (0.5 M H$_2$SO$_4$, pH 1) (black line) and alkaline solution (1 M KOH, pH 14). The $\eta_{10}$ values are indicated for each curve.

**XPS analysis of thermally and chemically treated MoSe$_2$ flake films**

**Figure S14** shows the XPS measurements on MoSe$_2$ flake films deposited on an Si substrate and annealed at different temperatures (600, 700 and 800 °C) in Ar/H$_2$ (90/10%) for 5 h. The Mo 3d XPS spectra confirm the progressive formation of elemental Mo (0) which correlates with the increase in the annealing temperature. In detail, the %c of the Mo (0) and the total Se is > 10% and < 20%, respectively, for annealing temperatures ≥ 700 °C. In these conditions, Mo (VI) is also observed with %c > 50%, which might be attributed to the subsequent oxidation of the elemental Mo under air exposure.[57]

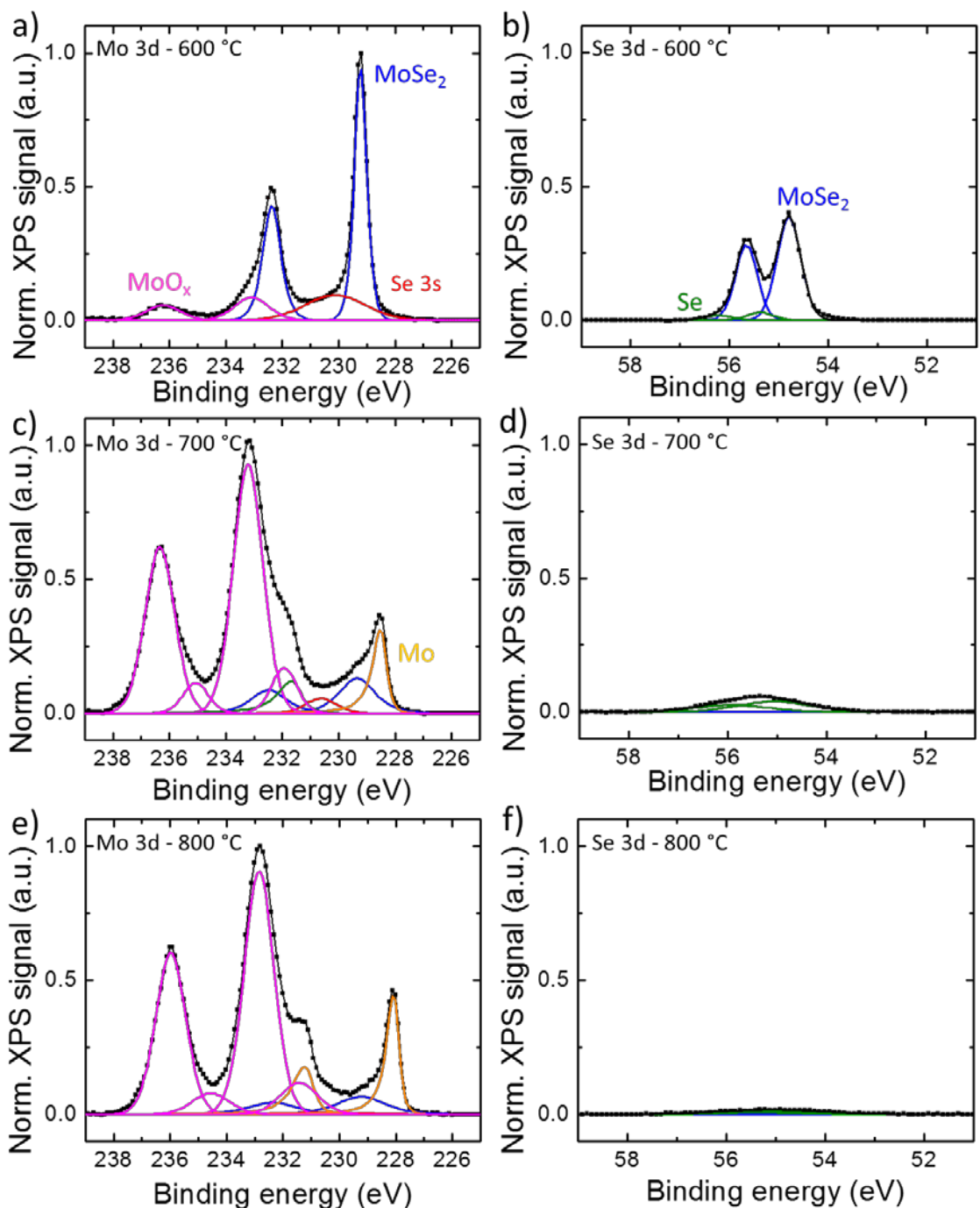

**Figure S14.** a) Mo 3d and b) Se 3d normalized XPS spectra for MoSe$_2$ flakes annealed at 600 °C in Ar/H$_2$ (90/10 %) for 5 h. c) Mo 3d and d) Se 3d normalized XPS spectra for MoSe$_2$ flakes annealed at 700 °C in Ar/H$_2$ (90/10 %) for 5 h. e) Mo 3d and f) Se 3d normalized XPS spectra for MoSe$_2$ flakes annealed at 800 °C in Ar/H$_2$ (90/10 %) for 5 h. Their deconvolution is also shown, displaying the bands ascribed to: MoSe$_2$ (blue curves); Se 3s (red curve), which overlaps with the Mo 3d XPS spectrum; oxidized species (MoO$_x$) (magenta curves); elemental Mo (orange curves); edge/elemental Se (green curves).

**Figure S15** shows the XPS spectra of the MoSe$_2$ flake films after chemical treatment, *i.e.,* after a12 h-chemical bathing in n-butyllithium. The results confirm the modification of the surface chemistry of $MoSe_2$. The spectra evidence the formation of different metallic phases, (*e.g.,* $MoO_x$ and Mo) and additional elemental atoms (Se and residual Li-species), which overlap with, and contribute to, the Mo 3d and Se 3d spectra of the $MoSe_2$ flakes (Li-species 1s XPS spectrum peaks between 50-60 eV), respectively. The $MoSe_2$-related XPS bands might be attributed to both the semiconducting (2H) and metallic (1T) phases.

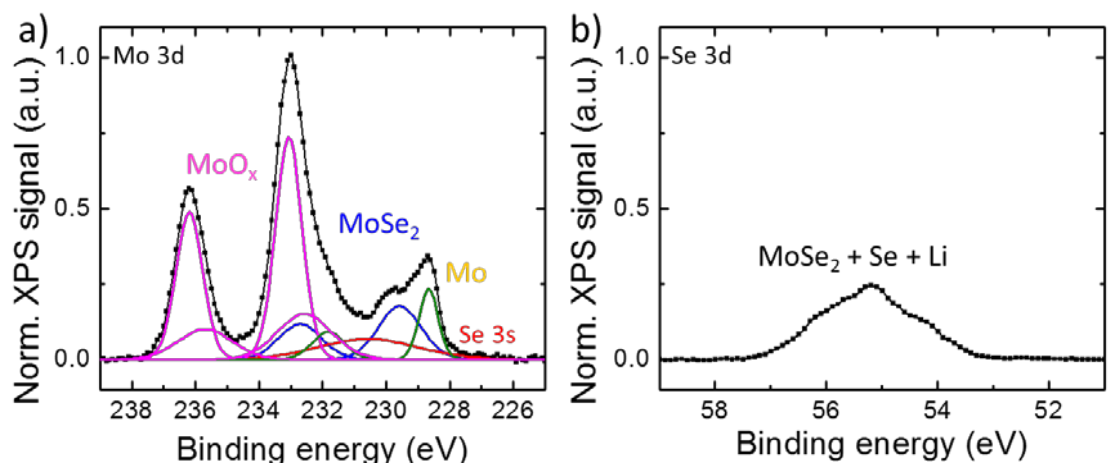

**Figure S15.** a) Mo 3d and b) Se 3d normalized XPS spectra for $MoSe_2$ flakes bathed in n-butyllithium for 12 h. The deconvolution of Mo 3d XPS spectrum is also shown, evidencing the band attributed to the: $MoSe_2$ (blue curves); Se 3s (red curve), which overlaps with the Mo 3d XPS spectrum; oxidized species ($MoO_x$) (magenta curves); elemental Mo (orange curves).

**Electrochemical characterization of glassy carbon/$MoSe_2$ electrode annealed at high temperatures in an $H_2$ environment**

**Figure S16** reports the polarization curves measured for the glassy carbon /$MoSe_2$ (GC/$MoSe_2$) annealed at 600, 700 and 800 °C in Ar/$H_2$ (90/10 %) for 5 h, in comparison to those obtained for the untreated electrode. The results confirm that the HER-electrocatalytic activity of the electrodes annealed at 600 and 700 °C is clearly enhanced with respect to that of the untreated electrode. In particular, the $\eta_{10}$ decreases from 0.34 V in the untreated electrode to 0.29 and 0.26 V in the electrodes annealed at 600 and 700 °C, respectively. A further increase in the temperature up to 800 °C causes a deterioration of the HER-electrocatalytic activity, whose $\eta_{10}$ (0.44 V) increases by 0.1 V with respect to that of the untreated electrode. Tafel slope values are also positively affected by the thermal treatment at 600 and 700 °C, for which they are 86 and 74 mV $dec^{-1}$, respectively. For the treatment at 800 °C, the lowest Tafel slope is observed (~144 mV dec$^{-1}$). The $j_0$ values increase with all the annealing temperatures, and are 19.09, 11.48 and 9.6 µA cm$^{-2}$ for 600, 700 and 800 °C, respectively. These results are explained by the thermo-induced texturization of the basal plane of the MoSe$_2$ flakes (see main text, Figure 8, and XPS analysis in the previous section, Figure S12). In detail, Se-vacancies, *i.e.,* HER-electrocatalytic sites, are formed due to H$_2$Se gas evolution during the thermal treatment in the H2 environment. This also causes an increase of the porosity of the electrodes, which is proved by the increase of their double-layer capacitance (Cdl), as measured by Electrochemical Impedance Spectroscopy (EIS) after their annealing (Figure S17) (0.66 mF cm$^{-2}$ for untreated GC/MoSe$_2$, 2.10, 3.10 and 5.11 mF cm$^{-2}$ for GC/MoSe2 annealed at 600 °C 700°C and 800 °C, respectively). However, at the highest annealing temperature of 800 °C, an excessive removal of Se could be detrimental for the HER-electrocatalytic activity due to the disappearance of the MoSe$_2$ phase, as demonstrate from $j_0$ values of the annealed electrodes, *i.e.,* 19, 11 and 10µA cm$^{-2}$ for 600, 700 and 800 °C (see main text).

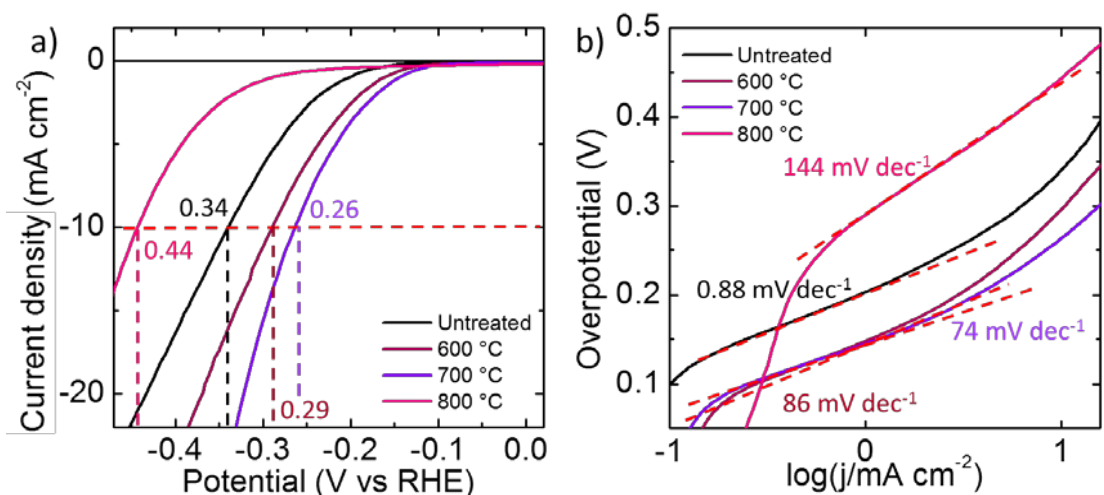

**Figure S16.** a) Polarization curves of untreated GC/MoSe$_2$ (solid black line) and GC/MoSe$_2$ annealed at 600 °C (solid purple line), 700 °C (solid violet line) and 800 °C (solid magenta line) in 0.5 M H$_2$SO$_4$. The η$_{10}$ values are also indicated for each curve. b) Tafel plots of untreated GC/MoSe$_2$ and GC/MoSe$_2$ annealed at 600 °C (solid purple line), 700 °C (solid violet line) and 800 °C (solid magenta line). Linear fits (dashed red lines) and the corresponding Tafel slope values are reported.

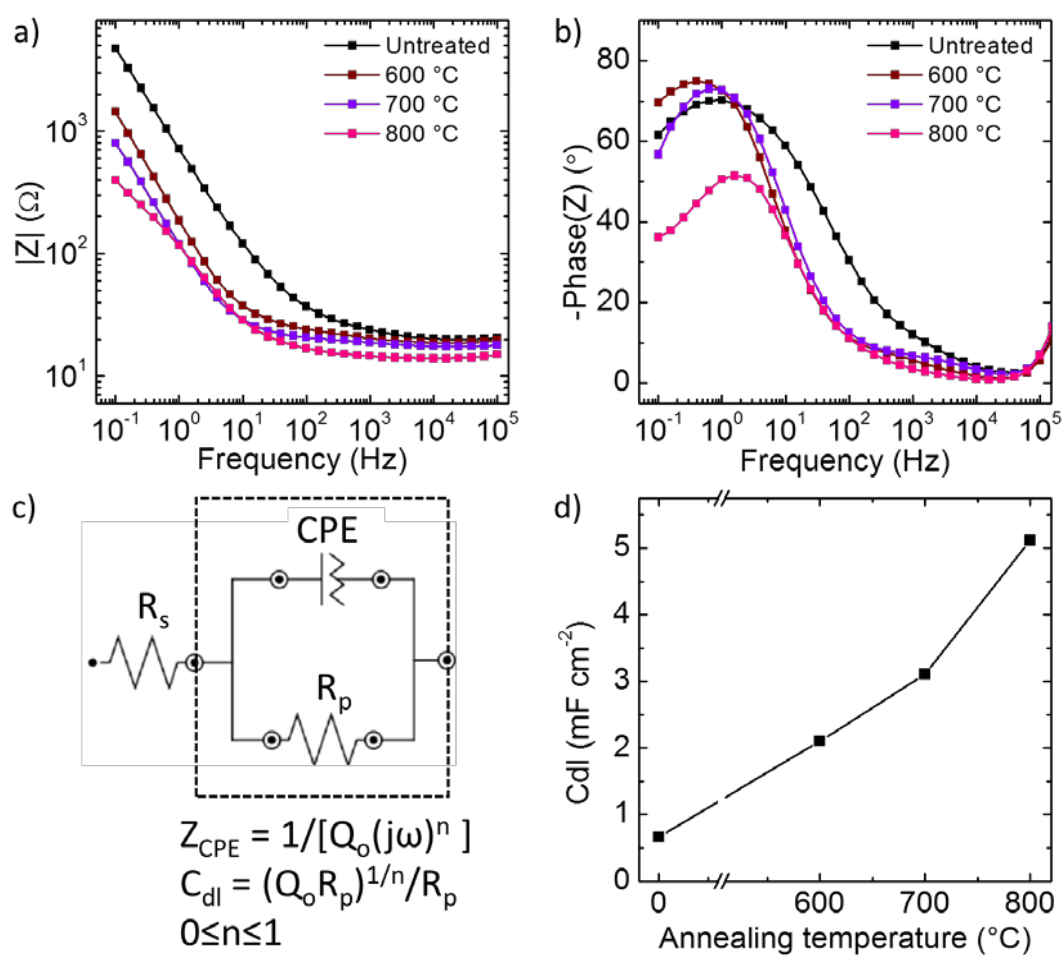

**Figure S17.** Electrochemical impedance bode plots for a) impedance magnitude (|Z|) and b) the negative impedance phase (-phase(Z) ) *vs.* frequency for GC/MoSe$_2$ before and after thermal treatment in H$_2$ atmosphere at 0.2 V *vs.* RHE (near the equilibrium potential of the electrodes) in 0.5 M H$_2$SO$_4$. c) Equivalent circuit used for extrapolating the double layer capacitance at the MoSe$_2$/electrolyte interface (C$_{dl}$). R$_s$ is the series resistance for the electrode and the electrolyte, while the parallel between the constant phase element (CPE) and the resistance R$_p$ is used for representing C$_{dl}$, in agreement with previous studies.[61,62] The equation for the impedance of CPE (Z$_{CPE}$) and the C$_{dl}$ are also reported. Q$_0$ and n ($0 \leq n \leq 1$) are frequency independent parameters. d) Plot for C$_{dl}$ *vs.* annealing temperature in H$_2$ atmosphere.

**Electrochemical stability of the MoSe$_2$-based heterostructures in operative HER-conditions**

**Figure S18** shows the chronoamperometry measurements (*j-t* curves) of the MoSe$_2$-based heterostructures over a period of 40 000 s (*i.e.,* > 11 h). For all the tested electrodes, a constant overpotential is applied in order to obtain starting current density of -30 mA cm$^{-2}$. The data show that the current density is retained for the untreated graphene/MoSe$_2$,

theSWCNTs/MoSe$_2$ and the SWCNTs/MoSe$_2$ annealed at 700 °C in an H$_2$ environment. Slight current density fluctuations might be caused by the consumption of H$^+$ or by the accumulation of H$_2$ bubbles on the electrode surface, which hinder the reaction.[58-60] For the SWCNTs/MoSe$_2$ that are chemically treated in n-butyllithium, the current density decreases by ~28%. The HER-electrocatalytic activity degradation might be due to the thermodynamically metastable nature of the 1T-phase, which could be converted back to its natural 2H-phase,[63-65] or to the dissolution of MoO$_x$ species in acid.[66-69]

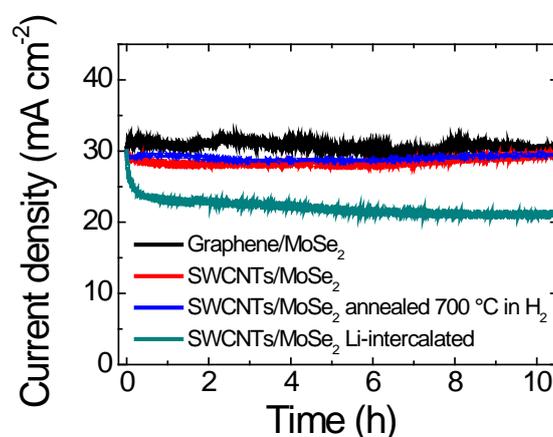

**Figure S18.** Chronoamperometry measurements (*j-t* curves) of the MoSe$_2$-based hetrostructures: graphene/MoSe$_2$ (black line), SWCNTs/MoSe$_2$ (red line), SWCNTs/MoSe$_2$ annealed at 700 °C in an H$_2$ environment (blue line) and SWCNTs/MoSe$_2$ Li-intercalated (*i.e.,* chemically treated in n-butyllithium) (cyan line) in 0.5 M H$_2$SO$_4$ (starting current density of -30 mA cm$^{-2}$)

**Figure S19** reports the XPS measurements of the surfaces of the GC/MoSe, graphene/MoSe$_2$ and SWCNT/MoSe$_2$ before and after the electrochemical stability tests. These data confirm that no significant changes occur for the MoSe$_2$-related bands both in the Mo 3d and Se 3d XPS spectra. The bands attributed to the edge/elemental Se are also preserved with their corresponding %c. This observation suggests that the active MoSe$_2$ phase is electrochemically stable during HER-operation. Interestingly, after HER, the MoO$_x$ species are not observed, suggesting their dissolution in acid,[66-69] thus not participating in the HER processes.

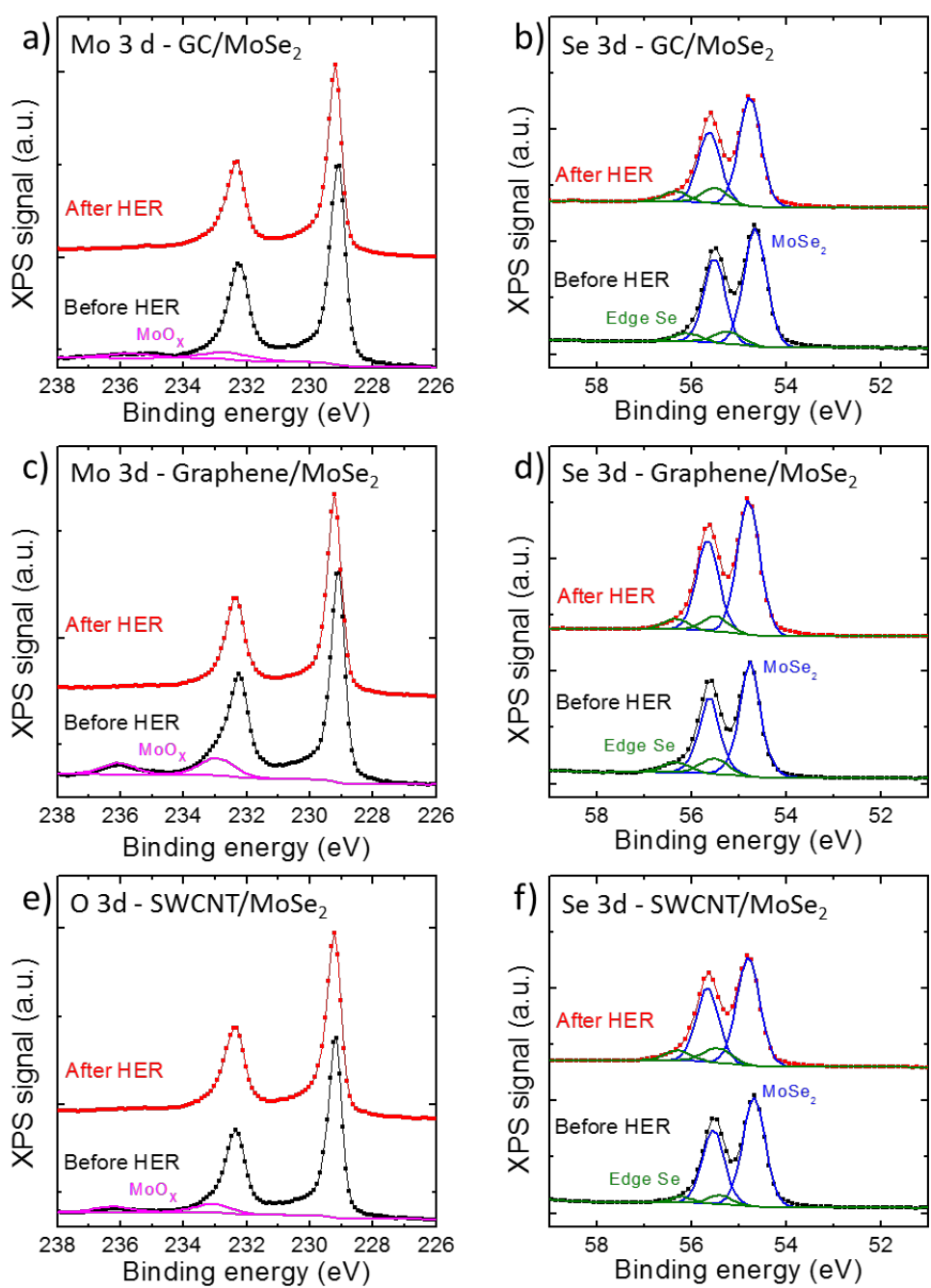

**Figure S19.** Mo 3d and Se 3d XPS spectra for a, b) GC/MoSe$_2$, c, d) graphene/MoSe$_2$ and e, f) SWCNT/MoSe2 before (black dotted lines) and after HER (red dotted lines). The bands attributed to MoO$_x$ species (magenta curves) in Mo 3d XPS spectra, and those attributed to the MoSe$_2$ (blues curves) and edge (elemental) Se (green curves) are evidenced.